\documentclass[12pt]{article}
\usepackage{amssymb}
\usepackage{amsmath}

\def\beq{\begin{equation}}\def\eeq{\end{equation}}
\def\bea{\begin{eqnarray}}\def\eea{\end{eqnarray}}

\begin{document}

\title{Causal set as a discretized phase spacetime}

\author{Roman Sverdlov
\\Raman Research Institute,
\\C.V. Raman Avenue, Sadashivanagar, Bangalore - 560080, India}
\date{May 3, 2010}
\maketitle

\begin{abstract}

The first goal of this paper is to show that discreteness, locality, and relativistic covariance can peacefully coexist if the ordinary spacetime (OST) is replaced with phase spacetime (PST) as a geometric background of a Poisson process, where PST is a spacetime generalization of a notion of phase space (this is a 7-dimensional version of the 8-dimensional structure proposed by Caianiello). Furthermore, Caianiello's idea of finite acceleration is implemented. After this is done, the paper then goes on to generalize the geometric notions obtained from the intuition of PST to a general discrete causal set, without any geometric background. It then takes advantage of the absence of lightcone singularity to attempt to tackle the definition of PST-like causal set; that is, a discrete system that approximates the geometrical properties that we would expect from continuum PST. Finally, the paper proceeds to introduce quantum field theory on a causal set, and shows that the locality gained by switching from OST to PST brings us one step closer to be able to treat quantum field theory on a causal set analytically rather than numerically. 

\end{abstract}

\noindent{\bf 1. Introduction}

It is well known that several approaches to quantum gravity rely on discretization of spacetime. If one is to perform that discretization in a relativistically covariant fashion, one would replace a continuous spacetime with Poisson distribution of points on a continuum background. However, as was discussed, for example, in \cite{Sorkin}, there are non-locality issues related to Minkowskian nature of space-time. In particular, it can be easily shown that the volume of a set of points whose Lorentzian distance to a given point $p$ is smaller than $\epsilon$ is infinite. Since in discrete scenario the distance is interpreted as a number of causal links, this implies that any given point has infinitely many other points to which it is connected by a single causal link. 

Furthermore, at least in a flat Minkowski space, for any given point p, and any reference frame, if we are to select a random point $q$ whose distance to point $p$ is less than $\epsilon$ then the difference in coordinates of these two points is arbitrarily large, with absolute certainty. This implies that if we are to introduce fields, such as scalar field, its variation between these two points might be arbitrarily large as well. Likewise, if we are to look at a curvature to analyze gravitational field, this can also vary by a very large amount between these two points. In continuum scenario this is not an issue because in order to take derivative of the field we are only concerned about the limit. Thus, the fact that $\epsilon$ is finite, however small it might be, makes the above issue irrelevant. In discrete case, however, we have no choice but to define "derivatives" in terms of finite distances, which makes the issue crucial.

Let us now analyze the root of the problem, in order to see how to address it. Our inability to distinguish different points $q$ that have the same Lorentzian distance, $\epsilon$, to point $p$ is related to the fact that one can perform Lorentz transformation that takes $q_1$ to $q_2$, while keeping $p$ fixed (here, the spacetime is assumed to be flat). Thus, if originally $q_1$ was on $t$-axis, while $q_2$ was near the light cone, and if our field distribution was "slowly varying" in the $pq_1$ frame, it would be seen as fast varying in $pq_2$ frame. 

This can also be explained as follows. In case of Newtonian physics, if we take two arbitrary frames, they will be moving with arbitrary high velocity relative to each other. The analogy of this statement for relativistic case is that two arbitrarily chosen reference frames move arbitrary close to the speed of light relative to each other. Thus, if we take two arbitrarily moving trains, due to Lorentz contraction, the length of each train will be arbitrary small in a reference frame of the other one. If a train is represented as a field, then each train will see another train as a $\delta$-function. The above Lorentz contraction argument continues to hold if a train was replaced with some smooth field distribution, which is the ultimate cause of the problem. 

Thus, the root of the problem is that the velocity of the trains relative to each other is arbitrary. This implies that the issue can be solved if we forbid the interaction between physical systems moving with very different velocities. This can be easily done by encorporation the idea of Caianiello (see, for example, \cite{Caianiello}) of adding the "velocity coordinates" to the spacial ones. In other words, we are "taking seriously" the concept of phase space and combining it with the concept of space time, to obtain the concept of "phase spacetime" (PST) as a basic background geometry. If the ordinary space time (OST) is represented by a d-dimensional manifold $\cal M$ (which, for convenience, can also be denoted by ${\cal M}_{OST} = \cal M$) PST is represented $2d-1$ dimensional manifold, ${\cal M}_{PST}$. This manifold consists of ordered pairs of the form $(x, v)$, where $x \in {\cal M}_{OST}$ and $v$ is a timelike tangent vector to ${\cal M}_{OST}$ at a point $x$, satisfying $g_{\mu \nu} v^{\mu} v^{\nu} =1$ (here, and throughout the paper, $(+1, -1, . . . , -1)$ metric convention is used). 

Just like two points that are "far away" in OST can't directly interact, neither can points that are "far away" in PST. In both cases, the interaction occurs through other points "in between", however fast. The only difference is that in terms of a projection onto OST coordinates, the distance between the points that can directly interact is very small, although it is still finite as a result of discretization that we postulated. On the other hand, the scale of direct interaction in the projection onto PST/OST might be \emph{either} small, \emph{or} large. It will be shown in section 4.4 that it is reasonable to assume that the scale corresponding to kinetic term of the Lagrangian is small, while the one corresponding to the interaction term is large. However, in both cases the scale is still finite. The finiteness of the interaction scale on PST/OST, however large it might be, allows us to enforce as small OST scale as we like.

A field $\phi$ is no longer a function of $x^{\mu}$ alone, but rather it is a function of both $x^{\mu}$ and $v^{\mu}$. In fact, if $v^{\mu}$ and $w^{\mu}$ are different vectors, there is no reason to expect $\phi (x^{\mu}, v^{\mu})$ and $\phi (x^{\mu}, w^{\mu})$ to be similar to each other, just like in case of OST we do not expect the fields at two different points that happened to have the same value of projection on one of the axes to be similar, either. Thus, it no longer makes sense to speak of $\phi (x)$, as the value can be anything, depending on the choice of $v$, which, also, can be anything. In section 4.4 we will find out, however, that if $\vert v- w \vert < \Lambda$, then interaction terms in the Lagrangian which correspond to, for example, $\phi^4$ in ordinary quantum field theory might couple $\phi (x, v)$ and $\phi (x, w)$, where $\Lambda$ is a large parameter corresponding to ultraviolet cutoff. However, in light of the small coupling constant, the values of a field at these two points are still fairly independent of each other.  

In such scenario, the non-locality problem can be addressed if we define continuity of $\phi$ to be a statement that $\phi (x, v) \approx \phi (x', v')$ whenever $(x, v) \approx (x', v')$, if $(x, v) \approx (x', v')$ is defined as follows:

1) $x$ and $x'$ are connected by unique geodesic, $\gamma (x, x')$

2) If $v''$ is a parallel transport of $v'$ from $x'$ to $x$ along $\gamma (x, x')$, then $g_{\mu \nu} (v''^{\mu} - v^{\mu})(v''^{\nu} - v^{\nu}) \approx 0$

3) In local geodesic coordinates around $x$ chosen in such a way that $v^0 =1$, and $v^k =0$, one has $x'^{\mu} \approx 0$ for any choice of $\mu$. 

It is important to notice that the topology on the PST is \emph{not} the product topology. To illustrate a point consider a flat spacetime. If $(x'^{\mu} - x^{\mu})(x'_{\mu} - x_{\mu}) <<1$, but $v^{\mu} (x'_{\mu} - x_{\mu}) =1$ (and $v=v'$), then $(x', v)$ is close to $(x, v)$ according to product topology, but not according to topology presented here. Since the product topology, by definition, does not use any information about $v$ in comparing $x$ and $x'$, one has no choice but to use Lorentzian distance to do the latter, which leads us right back to the infinities we were trying to avoid. On the other hand, the topology presented here makes explicit use of either $v$, or $v'$, or both, in order to select a reference frame in which to compare $x$ and $x'$. This distinguishes it from product topology, and also allows us to make sure that neighborhood of each point is finite; and in fact, can be made arbitrary small, as expected.  

In order to illustrate why the above does not violate Lorentz symmetry, lets go back to the case of OST. The only reason the notion of "Lagrangian density \emph{at a point}" does not violate translational symmetry is that action is defined to be the integral of Lagrangian density over \emph{all} possible points of the spacetime. Likewise, in PST case, if we were only interested in evaluating Lagrangian density at $(x_0, v_0)$, this would, of course, violate Lorentz symmetry \emph{along with} translational one. But the fact that we integrate over all possible $(x, v)$ to get the action is the reason that \emph{neither} translational symmetry \emph{nor} Lorentzian one were violated. 

Towards the end of the paper it will also be shown that this approach also makes it more realistic to be able to do analytic computations on a causal set, rather than numeric ones. To illustrate this, consider the causal set Lagrangians defined based on OST, in \cite{Proceedings} and \cite{paper5}. In the latter references, a neighborhood of any given point was defined in terms of Lorentzian distance, and thus was infinite. Thus, in order to define Lagrangian density, a sub-neighborhood of that neighborhood was selected. Each such sub-neighborhood, of course, had a corresponding "preferred" reference frame. 

In order for the choice of such frame not to violate relativity, the frame was defined as a function of the behavior of relevant fields. In particular, the frame was selected in a way that would minimize the observed fluctuations of fields (this, of course, was done in order to get rid of $\delta$-functions discussed earlier). Now, the procedure of selecting reference frame based on specific behavior of the field is, of course, very non-linear, and can not be represented in terms of perturbations to linearity, either. This is the ultimate reason why this approach does not allow one to go from Lagrangians to propagators, and ultimately forces one to use numeric methods.

On the other hand, according to the approach proposed in this paper, the quantum fields are functions on a PST, as opposed to OST. Since the neighborhood on PST is already finite, one does not have to select a sub-neighborhood. Since the latter is the only source of non-linearity, this modification opens the door for us to use ordinary perturbation theory techniques that are used in quantum field theory, and thus produce analytic results. 

\noindent{\bf 2. PST view of general causal set}

\noindent{\bf 2.1 OST view of a causal set: a review}

Even though the approach proposed on this paper replaces the OST with PST, it still borrows other aspects of the philosophy of causal set theory proposed by Rafael Sorkin. For that reason, it will be fruitful to review the original, OST-based, concept of a causal set, and then the PST concept will be introduced on the next section.

Suppose we have an OST, represented as a manifold ${\cal M} = {\cal M}_{OST}$, with a metric $g_{\mu \nu}$. Then the causal structure on $\cal M$ can be viewed as partial ordering, $\prec_{{\cal M}, g}$. For any $p$ and $q$, $p \prec_{{\cal M}, g} q$ if and only if $p$ and $q$ are connected by at least one timelike path $\gamma$ (timelike path is also sometimes referred to as causal path). In other words, $\gamma (\tau_1) = p$, $\gamma (\tau_2) = q$ and $g_{\mu \nu} d \gamma^{\mu} d \gamma^{\nu} >0$. This partial ordering is referred to as \emph{causal structure}.

Malement and Hawking have shown that there is a bijection between $g_{\mu \nu} / \vert g \vert$ and $\prec_{{\cal M}, g}$. Furthermore, in case of Poisson scattering of points, the uniform density of the scattering can be used to approximate $\vert g \vert$. This would imply that $\prec_{{\cal M}, g}$  approximates $g_{\mu \nu}$ itself. At the same time, on a microscopic scale, uncertainty principle implies that quantum fluctuations of gravitational field break down manifold structure, so $g_{\mu \nu}$ can no longer be used. A fundamental assumption of causal set theory is that partial order persists even on that scale, thus the latter is viewed as more fundamental than the metric, while metric is only an emergent manifestation of that partial order on a larger scale.

In light of this, geometrical quantities have to be defined on some more general (locally) finite set $S$, with a partial ordering $\prec_S$. For example, for any pair of points $p \prec_S q$, the distance between $p$ and $q$, $\tau_{S, \prec_S} (p,q)$ is defined as the product of some very small real number, $\tau_0$, with largest possible integer $n$ for which one can find a sequence $p \prec_S r_1 \prec_S . . . \prec_S r_{n-1} \prec_S q$. The number $\tau_0$ is interpreted as the length of a single link, or \emph{atomic scale}. It is introduced mainly in order to be able to carry out approximations in chapters 3 and 4, which depend on $\epsilon \delta << \delta$, where the distance scales defined by $\epsilon$ and $\delta$ are very small, but still large enough to include chains of large numbers of elements. This requires the length of each link to be much less than $1$, which is the purpose of the constant $\tau_0$.

Now suppose that there is some random mapping $f \colon S \rightarrow \cal M$. A \emph{pullback causal structure} on $S$ (see \cite{England}), which is denoted by $\prec_{f, g}$, is defined as follows: $p \prec_{f, g} q$ if and only if $f(p) \prec_{{\cal M}, g} f(q)$. It can be easily shown that in a local region of a manifold, where the geometry can be assumed to be locally flat and any pair of points is connected by unique geodesic, \emph{if} $\prec_S = \prec_{f, g}$ then $\tau_{{\cal M}, g} (f(p), f(q)) \approx k_d \tau_{S, \prec_S} (p, q)$ (see, for example, \cite{BrightwellGregory}). Here, $\tau_{{\cal M}, g} (f(p), f(q))$ stands for the length of the longest possible geodesic $\gamma$ in $\cal M$ connecting $f(p)$ and $f(q)$, the metric $g_{\mu \nu}$ is used to measure that length, and $k_d$ is some coefficient that depends on dimension, $d$, on a manifold.

While this point is controversial, this paper takes the point of view that $\tau_S (p,q)$ is exact rather than an approximation. After all, it is illogical to claim that a definition is not exact, since in this case one would need another, presumably exact, definition to compare it to. The latter would require a unique way of defining $\cal M$ into which $S$ is embedded, which would be a source of, presumably exact, quantity we are interested in. Even if this was possible, doing that would totally contradict the philosophy of causal set theory. In light of the fact that discrete causal structure is viewed as fundamental, while continuum is only something that is emergent on a large scale, the latter is an approximation for the former. This can be contrasted with the fact that in the case of Poisson distribution of points the discrete is an approximation for the continuous. This reinforces the idea of taking what used to be an approximation in continuum case, and make it exact for a causal set. 

In order to understand the way in which the definition and approximation reverse each other, a good example to think about is definition of trigonometric functions on a ring where the notion of limit, and therefore infinite series, is not defined. According to the philosophy of this paper, tangent of an element of a ring can be defined as a Taylor series up to some fixed constant, say $100$ (we will assume that multiplications and divisions by fixed real numbers are well defined). The expansion up to $100$ is, by definition, exact, and it continues to be exact even on the vicinity of $\pi /2$, including the point $\pi /2$ itself! On the other hand, the infinite Taylor series is no longer exact, but only an approximation; that approximation breaks down in vicinity of $\pi /2$. 

\noindent{\bf 2.2 Basic principles of PST discretization}

We would now like to rewrite the philosophy of the previous section, replacing the Poisson distribution in OST, with the one in PST. First, consider an $xy$ plane, where $x$ and $y$ are OST coordinates. A random scattering of points on $xy$ plane can \emph{not} be produced by first taking a random scattering on $x$ axis, and then, for each of the scattered values of $x$ take random scattering on $y$ axis. Likewise, in our case, since $x^{\mu}$ and $v^{\mu}$ are separate dimensions, it is \emph{not} correct to construct discretized PST by taking a union of random sprinklings on the tangent planes at randomly sprinkled points in OST. It is likewise not correct to view PST as a set of pairs of elements of OST since that, too, would result in PST fluctuations being the function of OST ones. Instead, we view PST as a single continuum manifold that was given from the start, and consider a single Poisson process involving scattering of points on PST itself. This means that if we wish to view the points in PST as vectors in OST, the probability of finding two vectors coinciding at a point is $0$. Besides, even if such pair of vectors did exist, it would be irrelevant, just like it is irrelevant if the $x$ coordinate of two scattered points in $xy$ plane happens to coincide. 

This raises a question: an element of PST is defined to be a vector in OST. In light of this, in order to define PST without first defining OST, we have to define a vector directly, without resorting to identifying it with a pair of points in OST. Of course, we would be able to do that if we are to assume the existence of a continuum geometrical background. But how can we rely on its existence, if that is something we ultimately intend to get rid of? We address this by recalling that in case of OST-based causal set theory, the definition of causal relations was also motivated by continuum geometry: two points are causally related if and only if one can go from $p$ to $q$ without going faster than the speed of light; the word "speed" requires the presence of geometry as well as metric in order to be well defined. So, in order to see whether our present geometric construction is "legal" or not, we have to outline what was done in case of causal set theory in OST, and see if we can do parallel steps for the case of a PST. For the discretization of an OST we did the following:

1) For a manifold $\cal M$, with metric $g_{\mu \nu}$, define a partial order $\prec_{{\cal M}, g}$. This requires the use of continuum geometry and metric. 

2) Define the notion of distances on ${\cal M}$, that is, $\tau_{{\cal M}, g} (p,q)$. Again, this requires continuum geometry and metric.

3) For arbitrary causal set $S$ with causal relation $\prec_S$, define $\tau_{S, \prec_S} (p,q)$. \emph{Note}: $S$ has nothing to do with $\cal M$, and, technically, $\tau_S$ has nothing to do with $\tau_{{\cal M}, g}$ either. Thus, no cheating was done.

4) For any function $f \colon S \rightarrow {\cal M}$ (which is interpreted as Poisson scattering of points on $\cal M$), and any metric $g_{\mu \nu}$ on $\cal M$, define a partial ordering $\prec_{f, g}$ on $S$ as follows: $p \prec_{f, g} q$ if and only if $f(p) \prec_{{\cal M}, g} f(q)$.  

5) Verify that, if $\prec_S = \prec_{f,g}$ then, with high enough probability, $\tau_{{\cal M}, g} (f(p), f(q)) \approx k_d\tau_S (p, q)$ for some dimension-dependent parameter $k_d$. This step is what we keep in mind while doing parts 3 and 4. However, since we didn't formally say it, no cheating has occurred.

Now, in light of the above scheme, geometry can be used to do parts 1 and 2. As long as one is allowed to use continuum geometry, one is allowed to use as much of it as necessary, including postulating tangent vectors to the manifold. Now, if we denote \emph{continuum} OST manifold by ${\cal M}_{OST} = \cal M$ and corresponding \emph{continuum} PST one by ${\cal M}_{PST}$, then the above 5 parts can be repeated, replacing ${\cal M}_{OST}$ by ${\cal M}_{PST}$ (and likewise replacing $f \colon S \rightarrow {\cal M}_{OST}$ with $f \colon S \rightarrow {\cal M}_{PST}$). The fact that continuum geometry was used in defining ${\cal M}_{PST}$ based on $\cal M$ has no impact on that statement. In fact, as you will shortly see, in order to define causal relations in PST even more advanced smooth geometry will be used, including geodesics, as well as their relative velocity at junction points. All of this will be "legal" because it will be understood that whenever such constructions are done, we are dealing with parts 1 and 2. Despite all this, the end result will not be continuum-based, thanks to parts 3 and 4.

 Furthermore, despite the fact that in parts 1 and 2 the ${\cal M}_{OST}$ was used to construct, and define geometry on, ${\cal M}_{PST}$, we still view ${\cal M}_{PST}$ as fundamental. The reason for this is that parts 1 and 2 are not an official part of the definition of a causal set; they are only a motivation. The verification of part 3 with parts 4 and 5 is independent of the \emph{way} in which the continuum was obtained; in other words, it is independent of parts 1 and 2. On the other hand, in light of differences in commutation properties of OST versus PST/OST, a special, non-trivial, construction will be invented in order to define OST as a subset of PST (which is to be contrasted with the way we started from OST to obtain PST in continuum case). This will be done through defining different geometrical quantities, such as "relative velocity" and "displacement" on PST causal set. Thus, two elements of PST causal set with non-zero relative velocity but zero displacement will be interpreted as two OST vectors at the same (OST) point. However, both "relative velocity" and "displacement" will be viewed as distance-like parameters, albeit with different properties. Furthermore, it will be understood that the discretized version of their commutation properties only applies to a small class of causal sets, that is referred to as PST-like. This is similar to the situation in causal set theory based on OST where manifold properties, such as approximate local validity of Pythagorean theorem, only apply to a small subset of causal sets, referred to as \emph{manifold-like} (or, in the language of this paper, \emph{OST-like}), even though the entire theory (including definition of length) is \emph{officially} applied to \emph{all} causal sets, while its secret aim is to get expected results \emph{on that small class}. 

Nevertheless, \emph{we are not completely done with parts 1 and 2}, so in much of the ramainder of the paper, we will use these parts in order to define the above parameters. Thus, while \emph{officially} we will view OST as a subset of PST, we will \emph{secretly} continue to view PST as an extension of OST while motivating our definitions. 

\noindent{\bf 2.3 Causal structure on PST}

Let us now move to the actual definition of causal relation on scattering of points on a PST. As was stressed in the introduction, our theory is based on the assertion that there is a fixed scale in which direct interactions are allowed. That scale is small in its projection onto OST, but it can be \emph{either} small \emph{or} large in terms of its projection on $PST/OST$. In order to be able to accomodate both, the geometry itself should have small scale, both on OST and on PST/OST. Moreover, since, as stated earlier, we do not separate these two sets, and scattering of points only respects the structure of \emph{entire} PST, we can simply say that the geometry has to be local, up to very small scale, on PST.  

In light of the philosophy of OST causal set theory, the geometry can be reproduced through causal relations, and causal relations are possible trajectories of photons. If we adopt that philosophy into PST causal set theory, we would conclude that we need photons to have velocity $v<c$ in order to "pass" by an element $(x, v)$. Furthermore, they have to have finite acceleration, in order to be able to "probe" the small $PST/OST$ scale (this, again, was originally proposed by Caianiello for continuum case -- see \cite{Caianiello}) . According to this model, the photons are undergoing the random walk in PST. In Newtonian case, the random walk in phase space amounts to velocity being random; and, most "random" velocities are arbitrary large. In relativistic case, the statement that most velocities are arbitrary large translates into a statement that most velocities are arbitrary close to $c=1$, which is a consequence of non-compactness of Lorentz group. By making the time scale of the random walk very small (but finite), we would see that after small, but finite, period of time the photon will \emph{almost} reach a velocity $c=1$, as desired.

As an important aside, we have to note that, as was mentioned in the introduction, and expended upon in section 4.4, the so-called "interaction" terms in the Lagrangian couple $\phi (x, v)$ and $\phi (x, w)$, where $\vert v - w \vert < \Lambda$, and $\Lambda$ is very large. Thus, the "interactions" that is used to probe small scales in PST are limitted to kinetic terms in the Lagrangian. The random walk in PST is something that photon does \emph{by itself}, without any interaction with matter fields. Depending upon the philosophy of a reader, one might say that while photon's interaction \emph{with matter} involves large momentum exchange, its interaction \emph{with vacuum} involves a very small one. This, of course, hints at the idea of vacuum energy. 

Let us now go back to the mathematical description of random walk in PST. We start out at $(x_1, v_1)$. Then, we make a "jump" in velocity, to $(x_1, v_1'')$ (the reason $v_1''$ is used instead of $v_1'$ will be clear shortly). Then, we make a "jump" in position by traveling to $(x_2, v_2')$,  where $(x_1, v_1'')$ and $(x_2, v_2')$ are tangents to the same OST geodesic $\gamma_1$. Then, again, we make a "jump" in velocity by going from $(x_2, v_2')$ to $(x_2, v_2'')$. Then, again, we make a "jump" in position by traveling along OST geodesic $\gamma_2$ from $(x_2, v_2'')$ to $(x_3, v_3')$, and we keep going in the similar fashion. Thus, we generate a sequence of points $(x_k, v_k)$ and OST geodesics $\gamma_k$ (or, in Lorentz notation, points $(x^{\mu}_k, v^{\mu}_k)$ and OST geodesics $\gamma_k^{\mu})$, where $k$ goes from $1$ through $n$. These OST geodesics are viewed to be differentiable functions $\gamma_k \colon \mathbb{R} \rightarrow \cal{M}_{OST}$, where $\cal M$ is OST manifold, satisfying 
\beq \frac{d^2 \gamma_k^{\mu}}{d \tau^2} = \Gamma^{\mu}_{\rho \sigma} \frac{d \gamma_k^{\rho}}{d \tau} \frac{d \gamma_k^{\sigma}}{d \tau} \; ; \; \gamma_k (0) = x_k \; ; \; \gamma_k (\tau_k) = x_{k+1} \eeq
such that the change of velocity upon passing each point is bounded by some parameter. Now, we notice that, in light of the fact that each $\gamma_k$ is parametrized from $0$ to $\tau_k$, we have $\gamma_k (\tau_k) = \gamma_{k+1} (0)$. Thus, the constraint on the change of velocity can be written as
\beq g_{\mu \nu} \frac{d \gamma_k^{\mu}}{d \tau} \vert_{\tau = \tau_k} \frac{d \gamma_{k+1}^{\nu}}{d \tau}\vert_{\tau = 0} < 1+ \epsilon \eeq
where the norm $1$ is assumed:
\beq g_{\mu \nu} v^{\mu} v^{\nu} = g_{\mu \nu} \frac{d \gamma_k^{\mu}}{d \tau} \frac{d \gamma^k_{\nu}}{d \tau} =1 \eeq
Now, if left as is, this would still have a lightcone problem, with $c$-lightcone being replaced with $v$-lightcone. This problem can be avoided by imposing additional constraint, that states that Lorentzian distance between $x_k$ and $x_{k+1}$ is smaller than $\delta$. In combination with norm 1 condition above, the distance constraint can be expressed as
\beq \tau_k < \delta \eeq
Since Lorentzian distance is $c$-based rather than $v$-based, most of $v$-based "light cone" is far away in Lorentzian sense. Equivalently, the criteria of two vectors being next to each other in a chain combines the $v$-lightcone with $c$-lightcone. It is easy to see that the combination of these two lightcones uniquely defines "preferred frame" (which points in a direction of the OST vector, corresponding to PST element of interest), and thus one has means of defining local region in that frame. 

We are now ready to define a causal structure on the scattering of points in PST. Two points $(y, u)$ and $(z, w)$ are causally related if there is at least one integer $n \in \mathbb{N}$ for which the above chain can be constructed, with $(y, u) = (x_1, v_1)$ and $(z, w) = (x_n, v_n)$ (this ammounts to replacing $\prec_{{\cal M}, g}$ with $\prec_{{\cal M}, f, g}$ since now it is a function of the scattering $f \colon S \rightarrow \cal M$, but this does not affect further steps, nor the conclusion of that argument). Thus, while the Lorentzian distance between two neighboring elements of a chain is bounded above, the distance between two arbitrary causally related elements is not. Incidentally, this borrows some aspects of graph theory advanced by Krugly, in that the original links (which are viewed to be edges of a graph) are non-transitive, while the (transitive) causal structure is defined by chains constructed out of these links. However, taking advantage of this extra structure is beyond the scope of this paper. 

The above discussion sums up in the following definition of a pullback causal structure $\prec_{f, g}$ for scattering $f \colon S \rightarrow {\cal M}_{PST}$ and the metric $g_{\mu \nu}$ on $\cal M$: 

\textbf{Definition}: Let $S$ be a set and let $f \colon S \rightarrow {\cal M}_{PST}$ be a random scattering on ${\cal M}_{PST}$. Then the \emph{pullback causal structure} on $S$ corresponding to $f$, which is denoted as $\prec_{f, g}$, is defined as follows: if $p$ and $q$ are elements of $S$, then $p \prec_f q$ if and only if one can find a sequence of other elements of $S$, $r_1, . . . , r_{n-1}$, and a sequence of differentiable functions $\gamma_1, . . . , \gamma_n \colon \mathbb{R} \rightarrow \cal M$ which satisfy the following conditions: 

a) $g_{\mu \nu} \frac{d \gamma_k^{\mu}}{d \tau} \frac{d \gamma_k^{\nu}}{d \tau} =1$

b) $\frac{d^2 \gamma_k^{\mu}}{d \tau^2} = \Gamma^{\mu}_{\rho \sigma} \frac{d \gamma_k^{\rho}}{d \tau} \frac{d \gamma_k^{\sigma}}{d \tau}$

c) $\gamma_k^{\mu} (0) = x_k^{\mu}$ and $\gamma_k^{\mu} (\tau_k) = x^{\mu}_{k+1}$, where $\tau_k < \delta$

d) $g_{\mu \nu} \frac{d \gamma_k^{\mu}}{d \tau} \vert_{\tau = \tau_k} \frac{d \gamma_{k+1}^{\nu}}{d \tau}\vert_{\tau = 0} < 1+ \epsilon$

e) $f(p) = \Big( \gamma_1 (0), \frac{d \gamma_1}{d \tau} \vert_{\tau =0} \Big)$ ; $f(p) = \Big( \gamma_n (\tau_n), \frac{d \gamma_n}{d \tau} \vert_{\tau =\tau_n} \Big)$

\noindent{\bf 2.4 Definition of Geodesic}

NOTE: from here on, the expression \emph{vector} means the same thing as "element of PST". This is similar to the expression "vector on a vector space". 

In the remainder of this chapter we will attempt to define parallel transports, rotations, distances and angles, which would give us all of the necessary geometry to be able to define PST geometry in chapter 3 and quantum field theory in Chapter 4. 

Most of the constructions that will be used rely on the notion of geodesics in discrete, PST-based causal set. In light of the fact that $f$-image of each element of $S$ (which, sloppily, will be identified with element itself from now on) can be viewed as a vector in OST, we would expect that each $p \in S$ determines unique geodesic passing through $p$. We have previously mentioned that causal structure on PST is motivated by a toy model of classical photons undergoing bounded-above changes in velocity in random walk fashion. In this case, the geodesic can be viewed as a "more likely" path of such photons.

However, while geodesic might be more likely than other paths, it is certainly not likely that the photon will follow that geodesic, or any other well defined path for that matter In fact, the application of random walk to flat spacetime tells us that the photon will move arbitrary close to $c=1$ with respect to any given reference frame, after enough time has passed, while geodesics have an arbitrary constant velocity $v<c$. Thus, we would like to say that geodesic is "the least un-likely path, while still very un-likely". More precisely, if an element $q$ is a distance $\tau_0$ away from the element $q$ on the geodesic passing through $p$, the probability of actually reaching $q$ after proper time $\tau_0$ is very small. Nevertheless, that same probability is much larger than the probability of reaching any other fixed element, $r$, after that same time $\tau$.

The discrete version of the path of fixed proper length is a sequence of points $p \prec^* r_1 \prec^* . . . \prec^* r_{n-1} \prec^* q$ with fixed $n$. Here, $\prec^*$ is a direct link; that is, $a \prec^* b$ if and only if $a \prec b$, and there is no $c$ satisfying $a \prec c \prec b$. Thus, $q$ is part of \emph{geodesic} passing through $p$ if there is at least one integer $n$ for which $q$ is connected to $p$ by more of the above sequences than any other PST element. However, in light of discreteness, there can be more than one element $q$ that meets that description. Since we don't have any other means of differentiating between them, we will consider a set of all such elements, for any fixed $n$. But, in the special case of an approximation to the continuum (which is what we are aiming for, in light of parts 1 and 2), these elements are so "close" to each other that they "look" like a single element (in PST terminology we have several points so close to each other that they "blur" into a single point; in OST terminology, we have few vector that are "very close" to each other and point in "very similar" direction that they "blur" into a single vector). So, even though, strictly speakng, it is a set, it "looks like" a translation of a single point $p$ by a fixed distance along geodeics. In light of this intuition, such subset of $S$ will be referred to as either \emph{past geodesic translation} or \emph{future geodesic translation} of an element $p$, depending on whether it is before or after $p$. It is strictly a translation along the direction of vector itself, and is not to be confused with a more general parallel translation, that simply happened to be time-like. This brings us to the following definition: 

\textbf{Definition}: Let $p$ be an element of $S$ and let $n>0$ be positive integer. Then \emph{future geodesic translation} of $p$ by $n$ is a set $G_n (p) \subset S$ consisting of elements $q$ with largest possible number choices of sequences $p \prec^* r_1 \prec^* . . . \prec^* r_{n-1} \prec^*q$. Furthermore, the \emph{past geodesic translation} of $p$ by $n$ is a set $G_{-n} (p) \subset S$, consisting of elements $q'$ with largest possible number of choices of sequences $q' \prec^* s_1 \prec^* . . . \prec^* s_{n-1} \prec^* p$. Finally, $G_0 (p)$ is a one element set, containing $p$ as its only element: $G_0 (p) = \{p \}$. 

Thus, we note that $G_m (p)$ is a set of elements that have coordinate $t=m$ in the reference frame defined by $p$. However, there are a number of features that one would not expect from ordinary geometry: 

1) The set $G_m (p)$ can contain more than one element, as stated earlier.

2) It is possible that the same element $q$ is an element of both $G_m (p)$ and $G_n (p)$ for different $m$ and $n$, as long as they are both non-zero and have the same sign.  

3) If $m<n$, $q_1 \in G_m (p)$, and $q_2 \in G_n (p)$, this does NOT mean $q_1 \prec q_2$, unless either $m$ and $n$ have opposite signs, or both of them are equal to $0$. In fact, it is even possible that $q_2 \prec q_1$.

Now, a geodesic passing through $p$, which is also referred to as \emph{trajectory of p}, is simply a union of all possible future and past translations of $p$: 

\textbf{Definition}: Suppose $p$ is an element of $S$. Then the \emph{trajectory} of $p$, or, equivalently, a \emph{geodesic} passing through $p$, is $G(p) \subset S$ defined as 
\beq G(p) = \bigcup_{n \in \mathbb{Z}} G_n (p) \eeq
The \emph{past} and \emph{future} trajectories of $p$ are $G_{-} (p)$ and $G_{+} (p)$, respectively, and are defined as 
\beq G_{-} (p) = G \cap J^{-} (p) \; ; \; G_{+} (p) = G \cap J^{+} (p) \eeq
where
\beq J^{-} (p) = \{ q \in G(p) \vert q \prec p \} \; ; \; J^{+} (p) = \{ q \in G(p) \vert q \succ p \} \eeq
In light of the fact that $G_0 (p)$ is one-element set while $G_m (p)$ is not, the geodesic has some "thickness" and it gets "thin" at $p$. This implies that, even if $q$ is an element of $G(p)$, $G(p)$ and $G(q)$ do \emph{not} coincide. 

\noindent{\bf 2.5 OST coordinates relative to an element of PST}

Since every PST element $p$, if viewed as an OST vector, determines a reference frame, it should be possible to separately define space and time displacements of some other element, $q$, relative to $p$. If photons were moving with speed $c$, we would imagine a photon being emitted at some element $r \in G_{m} (p)$, reaching $q$ and then after being re-emitted from $q$, being absorbed at $s \in G_{n} (p)$. Then, assuming $c=1$, the time and space displacements of $q$ are $(n+m)/2$ and $(n-m)/2$, respectively. 

I claim that we can continue to do the same thing in our case, despite the fact that speed of light is \emph{not} constant. After all, if the density of the scattering of elements on a PST is high enough, the photons will reach near-lightlike velocity within a very short time, with probability close to $1$. This means that the definition of the space and time displacements will approximate what we are used to on the scale much larger than the atomic time interval. 

Furthermore, as was discussed with the example of trigonometric functions, there is no such thing as a definition not being exact. Thus, by definition, the proposed notions of space and time displacements are exact, on all scales. On the other hand, since velocity was never defined to be a time derivative of position, but rather it was viewed as a separate coordinate, the latter can, in fact, break down. Thus, on the small scale we have
\beq v_r < \frac{dr}{dt} =1 \eeq
The above does \emph{not} refer to any effects of projections. In fact, even in $1+1$ dimensional case, what we used to view as a definition of velocity breaks down on a small scale. This is strictly a small scale effect, specific to this theory.

It should also be mentioned that the above construction relies on the fact that the element of interest is outside of the trajectory of $p$. After all, if the trajectory happened to have thickness of a single element (although it doesn't have to be) then the only way of applying the above construction to the element on the above trajectory would be for that element to send a photon to itself. This is not allowed since, conventionally, elements of a causal set are not causally related to themselves. For that reason, we have to separately define, by hand, that if $q \in G(p)$ then its space displacement is $0$.

Now defining the time displacement of $q$ is more complicated since it can be an element of both $G_m (p)$ and $G_n (p)$ for different $m$ and $n$. We would like to choose a definition of time displacement in a way that is the most consistent with the way we treat elements outside of $G(p)$. It is easy to see that this can be done by defining time displacement to be an average of the largest and smallest possible $m$-s satisfying $q \in G_m (p)$. 

Finally, if the space time curvature is large enough, it is possible that $G(p)$ "accelerates away" from some elements $s$ outside of $G(p)$ so fast that the imaginary photon emitted from $s$ would never "catch up". In such situation, we will formally define the spacial distance as $+ \infty$ and the time separation is $+ \infty$ if the lightcone of $s$ intersects $G_{-} (p)$, $- \infty$ if the lightcone of $s$ intersects $G_{+} (p)$, and $i \infty$ if neither intersections occur.

Of course, it is understood that the apparent infinities are most likely the consequences of curvature rather than the actual infinite distances. But, as was discussed in the trigonometry analogy, every definition is automatically true. Thus, the above definition is "legal" as long as used consistently. Now, in light of the fact that both intended geometry and intended physics are local, the distances between far away objects are not important, which is why the above definition is acceptable. The only purpose of infinity is to formally rule out these elements when it comes to defining a "neighborhood" based on the distance or time being less than some constant $\delta$. 

The above discussion can be summarized in the following definition: 

\textbf{Definition}: Let $p$ and $q$ be elements of $S$ and suppose $q$ is \emph{not} part of trajectory of $p$. Also suppose that $J^- (q) \cap G(p)$ and $J^+ (q) \cap G(p)$ are both non-empty. Let $m$ be the largest possible integer for which $G_m (p) \cap J^- (q)$ is non-empty. Likewise, let $n$ is the smallest possible integer for which $G_n (p) \cap J^+ (q)$ is non-empty. Then the \emph{time displacement} of $q$ relative to $p$ is given by 
\beq t_p (q) = \frac{n-m}{2} \eeq
and \emph{space displacement} of $q$ relative to $p$ is given by 
\beq r_p (q) = \frac{n + m}{2} \eeq
On the other hand, if either $J^- (q) \cap G(p)$, or $J^+ (q) \cap G(p)$ is empty, or both, then $r_p (q) = + \infty$. If $J^- (q) \cap G(p)$ is empty while $J^+ (q) \cap G(p)$ is not, then $t_p (q) = - \infty$. If $J^+ (q) \cap G(p)$ is empty while $J^- (q) \cap G(p)$ is not then $t_p (q) = + \infty$. Finally, if both sets are empty then $t_p (q) = i \infty$. Finally, if $q \in G (p)$, then $r_p (q) =0$ and $t_p (q)$ is the average between the largest and the smallest $n$ satisfying $q \in G_n (p)$. 

It is important to note that, as a result of the fact that it is conceivable to have $r_1 \prec r_2$, despite $r_1 \in G_m (p)$, $r_2 \in G_n (p)$ and $m>n$, one can also obtain negative space displacements. In particular, assume that $r_1 \prec s \prec r_2$ for some $s$ outside of $G(p)$. Let $M$ be the largest possible integer such that at least one element of $G_M (p)$ is to the past of $s$, and let $N$ be the smallest possible integer that at least one element of $G_N (p)$ is to the future of $s$. Since we already know that $m$ and $n$ satisfy these conditions, we know that $M \geq m$ and $N \leq n$. This, together with $m >n$ implies that $M >N$. Thus, $r_p (s) = (N- M)/2 <0$. 

However, in section 2.7, when the notion of \emph{PST-like} causal set will be defined, a vector-valued two point function $\vec{x} \colon S \times S \rightarrow \mathbb{R}^{d-1}$ will be postulated with a requirement that $\vert \vec{x} (p,q) \vert - \epsilon < r_p (q) < \vert \vec{x} (p,q) \vert + \epsilon$. This, together with the fact that $\vert \vec{x} \vert \geq 0$ will imply that, while $r_p (q)$ can still be negative, it can not be smaller than $- \epsilon$. 

Now that we have the definition of space and time coordinates with respect to a given element, let us now go back to the issue mentioned earlier, that the trajectory of the photon obeys $r=t$, despite the fact that its velocity is less than $1$. We will state it as a theorem, which we are now in a position to prove rigorously:

\textbf{Theorem}: Let $p$ and $q$ be two elements of a causal set $S$, and assume that $r_p (q)$ is finite. These two elements are causally related if and only if $\vert t_p (q) \vert \geq \vert r_p (q) \vert $. 

\textbf{Proof}: Since $r_p (q)$ is finite, we know that there are only two possibilities: either $q \in G(p)$, or else there is at least one element of $G_{-} (p)$ that is before $q$ and one element of $G_{+} (p)$ that is after $q$ (strictly speaking, the former case is a special case of the latter, but for rigor's sake they are treated separately). 

Lets start from the situation where $q \in G(p)$. In this case, by definition, $\vert r_p (q) \vert =0$. Since absolute value can not be negative, this automatically implies that $ \vert t_p (q) \vert \geq 0$, thus, $\vert t_p (q) \vert \geq r_p (q)$. At the same time, by definition of $G(p)$ there has to be a chain of points either of the form $p \prec r_1 \prec . . . \prec r_{n-1} \prec q$, or $q \prec r_1 \prec . . . \prec r_{n-1} \prec p$. Either of the two implies that $p$ and $q$ are causally related. Thus, in this special case, the equivalence between $p$ and $q$ being causally related and $\vert t_p (q) \vert \geq \vert r_p (q) \vert$ holds true: namely, both statements are true. 

Now lets move on to the situation where $q$ is outside the trajectory of $p$. Let $m$ be the greatest integer such that $q$ is after at least one element of $G_m (p)$ and let $n$ be the smallest integer such that $q$ is before at least one element of $G_n (p)$. As emphasized earlier, we do not know if $n<m$ or $n>m$. Thus, there are $3$ possibilities for either $m$ or $n$: each of them can be either negative, zero, or positive. Thus, total there are $3 \times 3 = 9$ possibilities. Some of them, however, will be shown as impossible. So I will show, case by case, that for every single one of these possibilities, one of the three things happens:

a) $\vert t_p (q) \vert \geq r_p (q)$ and $p$ and $q$ are causally related. 

b) $\vert t_p (q) \vert < r_p (q)$ and $p$ and $q$ are unrelated.

c) A contradiction is shown.

\textbf{Case 1:} $n<0$ and $m<0$. Then $r_p (q) = (n-m)/2$ (which can be both positive or negative) and $t_p (q) = (n+m)/2$. Then
\beq \vert t_p (q) \vert = - t_p (q) = \frac{-n -m}{2} = \frac{ \vert n \vert + \vert m \vert }{2} \eeq
and
\beq r_p (q) = \frac{n-m}{2} = \frac{ \vert m \vert - \vert n \vert }{2} \eeq
This implies that $\vert r_p (q) \vert < \vert t_p (q) \vert$, regardless of how $n$ and $m$ compare. 

Now, the way $n$ is defined tells us that there is $s \in G_n (p)$ satisfying $s \succ q$. Now, since $n<0$, $s \in G_n (p)$ implies $s \prec p$. This, together with $s \succ q$ implies that $q \prec p$. So in this special case both $q \prec p$ and $\vert r_p (q) \vert \leq \vert t_p (q) \vert$ hold.

\textbf{Case 2:} $n>0$ and $m>0$. Then $r = (n-m)/2$ (and again it can be both positive and negative) and $t_p (q) = \frac{n+m}{2}$. The fact that $n$ and $m$ are both positive immediately tells us that $\vert r_p (q) \vert < \vert t_p (q) \vert$ regardless of how the values of $n$ and $m$ compare to each other.

Now, the way $m$ is defined tells us that there exists $s \in G_m (p)$ satisfying $s \prec q$. But the fact that $m>0$ implies that $s \succ p$. This, together with $s \prec q$ implies that $p \prec q$. Thus, in this special case, $\vert r_p (q) \vert \leq \vert t_p (q) \vert$ and $p \prec q$ both hold.

\textbf{Case 3:} $m<0$ and $n>0$. The fact that $m<0$ implies that there are no elements of $G_0 (p)$ that are before $q$. The fact that $n>0$ implies that there are no elements of $G_0 (p)$ that are after $q$. Thus, together, they imply that every single element of $G_0 (p)$ is unrelated to $q$. But, by definition, $G_0 (p) = \{ p \}$. Thus, $p$ is unrelated to $q$. 

On the other hand, we know that 
\beq r_p (q) = \frac{n-m}{2} = \frac{ \vert n \vert + \vert m \vert}{2} \eeq
and
\beq t_p (q) = \frac{n+m}{2} = \frac{ \vert n \vert - \vert m \vert }{2} \eeq
which tells us that $r_p (q) > t_p (q)$.  

\textbf{Case 4:} $n<0$ and $m>0$. By definition of $n$ and $m$, there exists $s_1 \in F_m (p)$ and $s_2 \in F_n (p)$ such that $s_1 \prec q$ and $s_2 \succ q$. Now, the fact that $m>0$ implies that $s_1 \succ p$, and the fact that $n<0$ implies that $s_2 \prec p$. Thus, we have $p \prec s_1 \prec q \prec s_2 \prec p$ which implies $p \prec p$ which is a contradiction.    
 
\textbf{Case 5:} $m=0$ and $n>0$. The fact that $m=0$ means that $q$ is after at least one element of $G_0 (p)$. But, by definition, the only element of $G_0 (p)$ is $p$ itself. Thus, $q \succ p$. Also, we know that $r_p(q) = \frac{n}{2} = t_p (q)$. Since $\leq$ includes $=$, this means that $\vert r_p (q) \vert \leq \vert t_p (q) \vert$ holds.  

\textbf{Case 6:} $m<0$ and $n=0$. The fact that $n=0$ means that $q$ is before at least one element of $G_0 (p)$. But, by definition, the only element of $G_0 (p)$ is $p$ itself. Thus, $q \prec p$. At the same time, $t_p (q) = \frac{m}{2}$ and $r_p (q) = - \frac{m}{2}$. Thus, $\vert t_p (q) \vert = \vert r_p (q) \vert$. Since $\leq$ includes $=$, $r_p (q) \leq t_p (q)$ holds. 

\textbf{Case 7:} $m=0$ and $n<0$. By definition, there exists $r \in G_n (p)$ such that $q \prec r$. But the fact that $n<0$ implies that $r \prec p$. By transitivity, this means $q \prec p$. But the fact that $m=0$ implies that $q$ is after at least one element of $G_0 (p)$. Since the only element of $G_0 (p)$ is $p$ itself, we have $q \succ p$, which contradicts $q \prec p$. 

\textbf{Case 8:} $n=0$ and $m>0$. By definition, there exists $r \in G_m (p)$ such that $q \succ r$. But the fact that $m>0$ implies that $r \succ p$. Thus, by transitivity, $q \succ p$. But the fact that $n=0$ implies that $q$ is before at least one element of $G_0 (p)$. Since the only element of $G_0 (p)$ is $p$ itself, we have $q \prec p$, which contradicts $q \succ p$.

\textbf{Case 9:} $m=n=0$. By definition, there exists $r \in G_0 (p)$ and $s \in G_0 (p)$ such that $r \prec q \prec s$. But the only element of $G_0 (p)$ is $p$ itself, which means $p \prec q \prec p$ which is a contradiction. QED. 

\noindent{\bf 2.6 PST/OST coordinates relative to a PST element}

We have thus finished outlining the basic properties of $r$ and $t$. However, remembering that we are dealing with PST rather than OST, we are not done yet. Our next step is to define velocities. This, however, is easy to do: if we would like to measure velocity of an element $q$ relative to an element $p$, we just have to analyze the behavior of $r_p (s)$ verses $t_p (s)$ for a flowing element $s \in G(q)$. 

Now, if $t_p (q)$ and $r_p (q)$ are small compared to the scale of the curvature, we can select large enough $m$ and $n$ so that, if one "looks" at the picture from $G_m (q)$ or $G_n (q)$ one would see $p$ and $q$ being very close to each other. Thus, the angular velocity of $q$ relative to $p$ is approximately $0$ everywhere except a very small segment of the trajectory that happens to be close to $p$. This allows us to estimate the absolute value of the velocity of $q$ relative to $p$ without worrying about its angular velocity: 

\textbf{Definition}: Let $p$ and $q$ be elements of $S$ and let $n>0$ be a positive integer. Then \emph{$n$-th velocity} of $q$ relative to $p$ is given by 
\beq v_{p; n} (q) = \frac{\sum_{s \in G_n (q)} r_p(s)}{n \sharp G_n (q)} \eeq
where $\sharp$ stands for number of elements in a set.
   
The above definition violates time-reversal symmetry since $n$, which is assumed to be fundamental constant of nature, is either positive or negative. This is not a big deal since if we insist on having time reversal symmetry, we can always replace $v_{p;n}$ with $(v_{p;n}(q)+v_{p; -n} (q))/2$. The only reason it wasn't done that way is simply that if it is not done now it can always be done later, but if it is done now, it would not be possible to undo it, since some of the information would be lost.

It has to be noted that no assumption of a presence of geometry or smooth behavior is made. Thus, it is conceivable that $v_{p; n+1} = 10^6 \times v_{p;n}$. In this case, still, by definition, the velocity is exactly $v_{p; n}$. In fact, even if we had a situation where $v_{p; n} =1$ and $v_{p ; k} = 10^6$ for all $k \neq n$, the velocity will still be $1$, by definition. Again, the trigonometric analogy is what motivates this assertion. 

It is possible to avoid having to take relatively large distance in definition of velocity, by replacing velocity with Lorentz factor $\gamma$ that, as usual, represents time dilation. The reason $v$ was chosen was that the goal of this construction is to make things look as intuitive as possible, while the relation between $\gamma$ and $v$ requires a little bit of proof. Nevertheless, since this sense of simplicity is subjective, one might argue in favor of $\gamma$ since that would not require a time displacement by a relatively large distance. So for that purpose, definition of $\gamma$ will be included, and then it will depend on the purpose or philosophy of further work to choose between these two definitions:

\textbf{Definition}: Let $p$ and $q$ be elements of $S$ and let $n>0$ be a positive integer. Then \emph{$n$-th Lorentz factor} of $q$ relative to $p$ is given by 
\beq \gamma_{p; n} (q) = \frac{\sum_{s \in G_n (p)} t_p(s)}{n \sharp G_n (q)} \eeq
where $\sharp$ stands for number of elements in a set.

Now, in order to describe the motion of $q$ relative to $p$ fully, we do need to define its angular velocity. In case of usual geometry it can be estimated as $\omega = \frac{v}{r} sin \; \theta$, where $\theta$ is an angle between radius and the direction of velocity. It can be further seen that $sin \; \theta = h/r$ where $h$ is the distance between $p$ and the closest vector to $p$ on the trajectory of $q$. Putting these together we obtain  

\textbf{Definition}: Let $p$ and $q$ be two elements of $S$ and let $n$ be an integer. Then, \emph{$n$-th angular velocity} of $q$ relative to $p$ is given by
\beq \omega_{p;n} (q) = \frac{v_{p;n} (q)}{(r_p (q))^2} min \{ r_p (s) \vert S \in G(q) \} \eeq

This immediately gives the definition of radial velocity: 

\textbf{Definition}: Let $p$ and $q$ be two elements of $S$ and let $n>0$ be positive integer. Then the \emph{n-th radial velocity} of $q$ relative to $p$ is given by 
\beq v^R_{p; n} (q) = \sqrt{v_{p; n} ^2 (q) - r_p^2 (q) \omega^2_{p;n} (q)} \eeq

\noindent{\bf 3. Axioms that make causal set PST-like}

NOTE 1: Throughout this chapter, the distances are going to be scaled by a small parameter $\tau_0$, so that they no longer have to be integers and the smallest distance is much less than $1$. This is done in order to make sure that, for small distance scales $\delta$, we have $\delta^2 << \delta$ which would allow us to define approximations.

NOTE 2: This chapter can be skipped without any effect on understanding of chapter 4. 

NOTE 3: In light of discreteness I didn't even attempt to prove that any given set of axioms I propose will formally imply a specific geometry. I do have an intuitive sense that these specific axioms do imply that, but it is up to verification by numeric simulations (which haven't been done yet) to see to what extend this is the case. 

\noindent{\bf 3.1 Why axioms?}

Now we are done defining all of the fundamental parameters. It is important to note that, even though geometry was used as part of a motivation, no formal assumptions about the geometry were made in actual definitions. Thus, in case of an arbitrary causal set, we can still formally apply the above definitions to get space and time displacements as well as radial and angular velocities of different elements with respect to each other. The definitions were designed in such a way that in a very special case of \emph{PST-like} causal set they \emph{happen} to approximate the notions that we are used to. But since \emph{most} causal sets are not PST-like, in a general case these notions have nothing to do with any geometry; their definition is purely algebraic, just like, say, a definition of prime number.

To emphasize this point, we can take any given partially ordered set and do two separate exercises: one exercise involves defining OST-based quantities, and the other one involves PST-based ones. Despite the fact that these two sets of definitions are based on completely different assumptions of the geometry of light cone (which leads to different commutation properties of underlying space), both exercises can still be formally carried out on the same partially ordered set; but, of course, it doesn't mean that on a larger scale a given set can resemble OST and PST at the same time. In fact, we can, formally, define PST-based quantities on discretized OST, and OST-based quantities on discretized PST (I don't mean projection on OST, but rather a definitions based on OST interpretation of the shape of the light cone, which is why it is physically a wrong thing to do for PST). In both cases the procedure can be \emph{formally} carried out; but the results would completely lack any physical meaning. 

Because of the examples just presented, it is important to understand that PST-\emph{based} causal set is not the same as PST-\emph{like} one (the same is true if we replace PST with OST). When we say that a causal set is either PST-\emph{based} or OST-\emph{based}, we are referring to the definitions of some so-called geometrical parameters that are defined in purely algebraic way. Since these definitions can, formally, be made for any partially ordered set, any such set, regardless of its structure, can be viewed as either PST-based or OST-based, as we wish. On the other hand, when we say that a given causal set is PST-\emph{like}, we are saying that \emph{if} we choose to view it as PST-based, \emph{then} we will find that the causal structure is consistent with the one we will get through scattering of points in PST (and, of course, the same statement is true if we replace "PST" with "OST"). Thus, a statement that a given causal set is PST-\emph{like} (or OST-\emph{like}) \emph{does} say quite a bit about the specific structure of the partial ordering. 

It is also important to notice that the dynamics introduced in Chapter 4 relies \emph{only} on a causal set being PST-based, and it does \emph{not} assume that it is PST-like. This implies that for arbitrary partially ordered set the dynamics will still work in well defined fashion but it will not resemble anything we see in a lab. At the same time in a very special case of PST-like causal set it will. Even though we target that "very special case", the theory doesn't formally rely on it. For this reason, Chapter 4 can be read independently of Chapter 3. 

This leads to the following question. Since the dynamics only relies on a causal set being PST-based, what is the explanation that the causal set in which we are living in is PST-like? This question is by no means new. In fact, the OST-version of this question was asked for decades. One way to ask this question is this: since gravitation is a field, and that field corresponds to causal structure, there has to be a dynamics that "forces" causal structure to look like a manifold on a large scale, but such dynamics was not described yet. Unfortunately, there are a lot more problems with viewing gravitational field as field. The main ones are the following:  

a) While OST based gravity is described in \cite{Proceedings}, I have not come up with PST based gravity yet (it should be pointed out that this is only due to the fact that PST theory is very new, so it is not a serious objection). 

b) OST-based gravity shares the same faults as the version of scalar field Lagrangian introduced in \cite{Proceedings} (see section 4.1 for the summary of the latter). 

c) Even if gravitational Lagrangian was available, it is not clear how it can be used since, once geometry is allowed to fluctuate, we no longer have the geometrical background in which to use Lagrangians. An attempt to adress this issue was made in \cite{paper7}, but its success is very controversial.  

d) Suppose we did have a consistent theory of gravity with Einstein's equation being its classical limit, which would imply, for example, zero curvature for vacuum. In this case it is still not clear whether the zero curvature implies the presence of manifold structure. For example, in the OST situation, where the curvature is defined as in \cite{Proceedings}, that assertion is not true. Of course, it is possible that the situation would be different in this regard in PST scenario. But since no work has yet been done in defining PST curvature, this is still a subject of future research. 

e) Even if the dynamics was under control and classical limit of part d implied smooth manifold, we still do not know the mechanism by which the geometry "collapses" onto one smooth manifold structure rather than another. Again, an attempt was made to adress this issue in \cite{paper7}, but that theory is controversial.

For the reasons presented above, the dynamics that would generate PST-like causal set is postponned for future research. However, there is still a \emph{temporary} explanation as to why our causal set is PST-like, which appeals to the idea originally introduced by David Hume (\cite{HumeCausation}), and which I proposed to apply to quantum field theory and causal sets in \cite{paper7}. First, we make the following observation: any kind of dynamical law, such as the most basic classical physics, is a result of repeated observations. One might argue that there is no cause and effect, but rather there is a pattern of observed \emph{correlations} that we later \emph{interprete} as cause and effect. Thus, a statement "B happened because of A" becomes a statement "the history where A happens and B doensn't is not allowed, while the history where A happens first and B happens next is; thus, since we know A have happened, the only allowed correlations include event B afterwords". From this point of view, we never explained the occurance of B in terms of cause and effect. Instead, we simply \emph{postulated} the criteria for "allowed" correlations, and then occurence of B became a logical (\emph{not} causal) consequence of that postulat, together with occurence of A. 

The implication of the above philosophy is that the explanation of something is simply recognizing the pattern. For instance "the sun is rising this morning because the sun is rizing \emph{every} morning" is a good example of explanation, even though I never used either Newtons or Keplers laws to explain why the sun rises every morning. Now,this philosophy is perfectly consistent with my \emph{eventual} use of Newton's laws to explain the behavior of the sun. After all, the latter explanation would amount to saying "sun does Y because \emph{every} physical object does Z", where Y amounts to rising every morning, while Z amounts to obeying the Newton's gravitation laws. On the other hand, the \emph{temporary} explanation we had prior to that is that sun does X (i.e. rises up this week) because it does Y (i.e. rizes up every day). So Y is bigger pattern than X, and Z is bigger pattern than Y. Thus, generalizing X into Y and generalizing Y into Z both fit the definition of "explanation". After all, an explanation amounts to showing how some smaller pattern can be viewed as part of bigger one. Since "smaller" doesn't mean "small", and "bigger" doen't mean "big", it is always possible to "explain the explanation" by showing how bigger pattern is part of even bigger one. In fact, since any theory is based on axioms, which, in principle, can further be explained by other axioms. Thus, there is no such thing as "ultimate explanation". 

In light of this philosophy, the explanation of why causal set is PST-like is similar to sun rizing up every morning, while the actual dynamics that generate such a causal set is similar to the Newton's law of gravity. Since we claim that it is okay to view "the sun rizes this morning because it rizes every morning" as a satifactory theory \emph{until} the law of gravity is discovered, we likewise claim that the explanation "the causal set we are living in is PST-like because every causal set is such" is also an okay explanation, until dynamics is introduced. 

Of course, we have all observed PST-like pattern from our everyday experience, so we are almost done; but not quite. After all, by PST we mean a continuum background, while causal set is discrete. So, our experience presumably teaches us a \emph{discretized} version of PST-like pattern. Thus, in order for that "observation" to count we need to come up with rigourous definition of what observation are we referring to. In case of sun rizing up every morning, in order for that "theory" to be complete, we have to define what we mean by the morning. For example, we can say that a morning is when our watch shows such and such time, and then define what we mean by well working watch. Thus, the definition of well working watch \emph{as opposed to} the law of gravity becomes a necessary component of the explanation why the sun will rise at a given time. Likewise, in the case of PST-like causal set, we have to define what we mean by PST-like; once we do, then we can simply postulate it as an "axiom" that the causal set has to be PST-like, and that axiom will be the explanation we are seeking.  

Finally, it is important to point out that axiomatization of OST has been dealt with, in a lot more rigorous way, in \cite{Robb} and \cite{Malement}. In this paper I can't possibly measure up to their standard since I am not dealing with continuum, hence I lack a mashinery to do rigourous proof. At the same time, thees papers have their own weakness: once the curvature is introduced, they become approximations, which break down in the vicinity of light cone. In light of the fact that we have addressed the lightcone problem by introducing PST, it might be of interest to see if axioms can be invented in order to define, with specified degree of accuracy, PST-like causal set, while accommodating curvature.  It is also possible that PST version of \cite{Robb} might be simpler than OST one since in case of a PST the timelike geodesics arises in a natural way, so one does not need to introduce intersection of timelike planes to define them. However, the axiomatization proposed in this paper is completely different from the one of \cite{Robb} and, therefore, the analysis of the approach of \cite{Robb} or its modifications is beyond the scope of this paper. As far as this paper is concerned, some short cuts were made which were not doable within a context of \cite{Robb}, such as appealing to the notion of radial, time, and velocity coordinates. This means that, as part of the future research, I can abandon these assumptions and still investigate if translating \cite{Robb} to PST scenario will allow the axiomatization without such shortcuts.

\noindent{\bf 3.2 The philosophy behind the choice of axioms}

Since we are trying to impliment \emph{local} PST-like topology rather than global one, we have to define the notion of locality. In light of continuum, the "local" neighborhood can be arbitrary small. In discrete scenario, on the other hand, one requires a certain constant that defines the size of generic local neighborhood. This constant can not be imposed on Lorentzian distance since, as discussed earlier, most points $q$ that have Lorentzian distance to $p$ smaller than $\epsilon$ are still arbitrary far from $p$ coordinate-wise. So, instead, the constraint is imposed on $t$, $r$ and $v$ coordinates discussed in the previous chapter. More precisely, I postulate a \emph{small} numbers $\delta$ and $h$, as well as a \emph{large} number $\Lambda$, and define the $(h, \delta, \Lambda)$ neighborhood of $p$ as a set of all elements $q$ satifying $\vert t_p (q) \vert < h$, $r_p (q) < \delta$, and $\vert v_p (q) \vert < \Lambda$. The reason $\Lambda$ is assumed to be large is that we do not have the curvature effects in velocity (i.e. PST/OST) directions while we do have them in position (OST) ones. On the other hand, the reason $\Lambda$ was introduced at all is that in a reference frame that moves very close to the speed of light, the length of a single causal link "stretches out" very far coordinate-wise. Thus, even if $\vert t_p (q) \vert$ and $r_p (q)$ are both small, this doesn't mean very much since their definitions involve "very long" links, and thus their error can, possibly, be of a magnitude of several thousands kilometers! Finally, for the sake of convenience of sparing one parameter, the two "small" parameters $h$ and $\delta$ will be assumed to be equal, which allows letter $h$ to be dropped.

It might seem that the fact that $\Lambda$ is finite (although large) violates the Lorentz covariance. As explained earlier, this violaltion of Lorentz covariance is similar to the violation of translational covariance by a small neighborhood that we are using to from OST-based quantum field theory that existed for decades. But, as stated previously, the philosophy of this paper is that, while both translational and rotational symmetry is violated by any given neighborhood, it will be restored after we consider all possible neighborhoods. In light of this philoophy, we are allowed to violate Lorentzian symmetry \emph{as much as we like} while defining the axioms for any given neighborhood. The two particular ways in which we insist on violating relativity is introducing finite $\Lambda$ and separating space and time coordinates. In fact, not only we can do these things, but in fact we should! After all, if we explicitly separate space and time coordinates and still get relativity at the end of the day, this will prove a point that relativity arises naturally as a consequence of this approach. On the other hand, if we didn't separate space and time, we would have been using minus signs in Lorentzian signature to combine them, and then one would argue that we put relativity by hand.

Now, the easiest way to argue that relativity can be restored after having separated space and time is to use the result of numerical studies done in \cite{BrightwellGregory} which shows that in case of the random scattering of points on Minkowskian space, there is a very close correlation between the longest chain of points and the Lorentzian distance. Thus, by using former instead of the latter, we no longer need to remember the minus sign of Minkowskian metric. However, we still have to remember the definition of a causal structure in order to be able to list the chains of points available. The causal structure is defined in terms of light cone, so we have to remember that the speed of light is constant. Furthermore, in order to define speed of light, we have to remember the definition of distances on spacelike hyperplane (for flat case). That definition relies on Pythagorean theorem. While it is true that in the previous chapter the distances were defined without Pythagorean theorem, one can not perform numerical studies on generalized spaces, including the one discussed there. 

This raises a question: should Pythagorean theorem be postulated, or should it be derived? On the one hand, the philosophy of the paper only demands that we refrain from postulating relativistic effects; in this respect it is okay to postulate Pythagorean theorem since it is not one of them. On the other hand, a Pythagorean theorem is a consequence of rotational symmetry, and the latter can be obtained through the commutation of Lorentz transformations. In light of this, one might argue that if we "don't know" Lorentz transformations, we shouldn't "know" Pythagorean theorem either. However, one might also argue that rotational symmetry is only an emergent notion that arises on a large scale. If we are to define the discretized version of that symmetry, we would have to introduce a lot of "small" constants in order to set up a rigourous criteria of what we mean by \emph{approximate} symmertry (in fact, even though we are not going to take the rout of defining commutation relations, we will still have to introduce constants in sec 3.4 through 3.6 while defining other "axioms" that are expected to hold "approximately true"). 

In light of this it is the matter of taste as to what approach is more natural. On the one hand, postulating Pythagorean theorem requires less axioms and less small constants, so it is more natural from the formal point of view. On the other hand, Pythagorean theorem can not be explained to the first grader, while the axioms and small constants through which it is derived can be; so in this respect they are more intuitive. In light of this controversy, I leave it to the reader to choose between sec 3.3 where Pythagorean theorem is postulated by hand and sec 3.4 through sec 3.6 where other axioms are postulated in order to derive it. In the latter case one can skip sec 3.3 since its results are not used in sec 3.4 through 3.6. 

In fact, if one is not happy with the philosophy of postulating emergent smooth geometry at all, one can skip the rest of this chapter altogether, and still read chapter 4. As mentioned earlier, According to philophy of this paper, the quantum field theory proposed in Chapter 4 is applicable to \emph{all} causal sets, not just PST-like ones. While it is true that a very special case of PST-like causal sets was used to \emph{motivate} the setup of quantum field theory, the latter is formally defined without an assumption of the structure of a causal set, and is meant to apply to all causal sets. Furthermore, when PST-like structure is assumed in motivational part, one is only assuming the Poisson scattering in continuum PST background; thus, neither of the axiomatization approaches are used. In light of this, Chapter 4 is completely independent of Chapter 3, and one might skip Chapter 3 entirely if one wishes to do so. 

\noindent{\bf 3.3 Defining PST-like causal set while making use of discretized Pythagorean theorem} 

As explained in the previous section, we are going to separate space and time coordinates, and impose Pythagorean theorem on the space ones. Thus, the OST coordinates can be represented by two-point functions $\vec{x} \colon S \times S \rightarrow \mathbb{R}^{d-1}$ and $t \colon S \times S \rightarrow \mathbb{R}$, that satisfy $r_p (q) \approx \vert \vec{x} (p,q) \vert$ and $t_p (q) \approx \vert t(p,q) \vert$. The PST/OST coordinates, on the other hand, are represented through the "velocity" function $v \colon S \times S \rightarrow \mathbb{R}^{d-1}$, satisfying $\vert \vec{v} (p,q) \vert \approx v_{p; n} (q)$. These are summarized in part b of the definition below. One has to remember to enforce transitivity, that is, if 
\beq \vec{v} (a, b) \approx \vec{v} (a, c) \approx \vec{v} (b,c) \approx \vec{0}\eeq
then  
\beq \vec{x} (a, c) \approx \vec{x} (a, b) + \vec{x} (b, c) \eeq
and 
\beq t(a, c) \approx t(a, b) + t(b, c) \eeq
This corresponds to part c of definition below. 

Now, the above "transitivity" relations only apply to the PST elements (or, equivalently, OST vectors) whose respective velocity relative to each other is close to $0$. In order to include other PST elements we have to impose a constraint that if in the frame of PST element $b$, the OST coordinates of PST element $c$ are close to $0$, then, in the frame of element $a$ the OST coordinates of PST elements $b$ and $c$ are nearly the same. This will allow us to use PST element $b$ whose velocity relative to PST element $a$ is small in order to assess the OST coordinates of $c$, whose velocity relative to $a$ might be large. However, it is important to notice that while the velocity of $c$ relative to $a$ might be large, it can't be too close to $1$ since, in this case, due to relativistic effects, the small shifts due to Poison scattering in one frame might look like very "long" in another one. So for this reason the approximate statement only applies when $v_b (c) < \Lambda$, where $\Lambda$ is some fixed real number that is very close, but smaller than, $1$. That statement is included, in a more explicit form, in part d of definition below. 

The expected Lorentzian distances statistically arise only if the density of scattering is uniform, so it is crucial to postulate the latter. In order to do that, we have to first define density, and then constrain it to a range between $\rho - \lambda$ to $\rho + \lambda$ for some specified $\lambda << \rho < 1$. While it is possible to define density in terms of Poisson distribution, $e^{-\rho} \rho^n /n!$, I choose not to do that since this equation can be derived analytically from much simpler definition: the density is the number of very small volume elements that contain at least one scattered point. Now, if we view point $p$ as a local origin, in order for theory to be transitive, we would like to define density in the $\delta$-neighborhood of $p$, rather than $p$ itself. In particular, we would like to select $t$, $\vec{x}$, and $\vec{v}$, satisfying $max (\vert t \vert, \vert \vec{x} \vert) < \delta$, and $\vert \vec{v} \vert < \Lambda$, where $0< \delta < \Lambda <1$, $\delta \approx 0$ and $\Lambda \approx 1$. We would then like to define the notion of probability of finding an element $q$ whose coordinates relative to $p$, $(t', \vec{x}', \vec{v}')$ satisfy $\vert t' - t \vert < \chi \delta$, $\vert \vec{x'} - \vec{x} \vert < \chi \delta$, and $\vert \vec{v'} - \vec{v} \vert < \chi \delta$. We can define this probability by simply counting the number of all such small regions and see how many of them actually contain an element of PST. 

We would formally accomplish this by defining $B_{\delta \chi}$ to be a set of quadruples $(p; t, \vec{x}, \vec{v})$ (where $max (\vert t \vert, \vert \vec{x} \vert) < \delta$ and $\vert \vec{v} \vert < \Lambda$), for which there is at least one element of $S$ with coordinates $(t', \vec{x}', \vec{v}')$ that differ from $(t, \vec{x}, \vec{v})$ by less than $\chi \delta$. We then put a restriction on the measure of such set, $\mu_1 < \mu B_{\delta} < \mu_2$. All of this will be done in part $f$ of the definition that follows:  

\textbf{Definition}: Let $d$ be an integer. Causal set $S$ is \emph{PST-like} up to $(d; \delta, \epsilon, \mu_1, \mu_2, n)$ if there are functions $\vec{x} \colon S \times S \rightarrow \mathbb{R}^{d-1}$, $\vec{u} \colon S \times S \rightarrow \mathbb{R}^{d-1}$ and $t \colon S \times S \rightarrow \mathbb{R}$ satisfying the following conditions:

a) $t (p, p) = 0$, $\vec{x} (p, p) = \vec{0}$, and $\vec{u} (p, p) = \vec{0}$

b) For any $p$ and $q$ such that $\vert t_p (q) \vert < \delta$, $\vert r_p (q) \vert < \delta$, and $\vert v_p (q) \vert < \delta$, the following is true:
\beq t_p (q) - \epsilon \delta < \vert t (p,q) \vert < t_p (q) + \epsilon \delta \eeq
\beq r_p (q) - \epsilon \delta < \vert \vec{x} (p,q) \vert < r_p (q) + \epsilon \delta \eeq
\beq v_{p; n} (q) - \epsilon \delta < \vert \vec{v} (p,q) \vert < v_{p;n} (q) + \epsilon \delta \eeq

c) If $p$, $q$ and $r$ are three elements and $max (\vert \vec{v}(p,q) \vert, \vert \vec{v}(p,s) \vert, \vert \vec{v}(q,s) \vert)< \delta$, then $\vert t(p,q) + t(q, s) - t(p,s) \vert < \epsilon \delta $ and $\vert r(p,q) + r(q, s) - r(p,s) \vert < \epsilon \delta$

d) If $max (\vert r_a (b) \vert, \vert t_a (b) \vert ) < \delta$, $max ( \vert t_b (c) \vert, \vert r_b (c) \vert) < \delta^2$ and $max (\vert v_a (b) \vert, \vert v_b (c) \vert < \Lambda$, we have $\vert \vec{x} (a, c) - \vec{x} (a, b) \vert < \epsilon \delta$ and $\vert t (a, c) - t (a, b) \vert < \epsilon \delta$

e) If $p$, $q$ and $s$ are three elements of $S$, $s \in G(q)$, $\vert r_p q \vert < \delta$, $\vert t_p q \vert < \delta$, and $\vert v_p q \vert < \Lambda$, then 
\beq \vert \vec{x} (p, s) - \vec{x} (p,q) - \vec{v} (p,q) t (q,r) \vert < \epsilon \delta \eeq

f) Let $B_{\delta}$ be a set of all quadruples $(p, t, \vec{x}, \vec{v})$ satisfying $max (\vert t \vert, \vert \vec{x} \vert)< \delta$ and $\vert \vec{v} \vert < \Lambda$ for which there exists at least one $q$ satisfying $\vert t(p,q) - t \vert < \delta^2$, $\vert \vec{x}(p,q) - \vec{x} \vert < \delta^2$, and $\vert \vec{u}(p,q)  - \vec{u} \vert < \delta^2$. The measure of the set $B_{\delta}$, defined as above, satisfies 
\beq \mu_1 < \mu B_{\delta} < \mu_2 \eeq

\noindent{\bf 3.4 Timelike coordinate parameters}

In light of the fact that PST element corresponds to timelike vector on OST, we claim that a local choice of $d$ elements of PST spans a local OST region. Furthermore, in light of the fact that velocities are defined in terms of OST parameters (time and distance), the above choice of $d$ PST elements can ultimately be used in identifying PST/OST coordinates as well, which will also be called \emph{velocity coordinates}. As was also true in OST case, it should be understood that, in light of discreteness, the point-wise definition of these statements is not true; thus, the notion of spanning can only be defined in statistical way. 

One of the fundamental building blocks of coming up with definition of spanning is the idea that the definition of relative velocity in section 2.6 allows us to define the notion of "co-moving" elements of PST (or, equivalently, "parallel" OST vectors) as the ones with relative velocity close to $0$. The fact that relative velocity was defined in non-local way, implies that the notion of co-moving vectors is equally non-local, and does not require the notion of parallel transport in order to be well defined (but in practice only the vectors within OST $\delta$-neighborhood of each other form coordinate system that has physical meaning, which means that the word "non-local" should be replaced with "quasi-local"). This choice was motivated by the fact that in order to define discretized parallel transport one would have to introduce a sequence of parallel elements such that each next element of the sequence is "very close" to the previous one. This requires one to be able to define a notion of "nearby" elements being parallel to each other. Now, if two elements are trully nearby, there might be too much random fluctuation to mess up the parallelism we are seeking. On the other hand, if the two elements we are interested in are separated by a distance of order $\delta$ then, even though $\delta$ is assumed to be small, there is enough intermediate elements for the statistical fluctuations to average out, which makes the notion of parallelism more reliable. Thus, ironically, the quasi-local notion of parallelism becomes more reliable than parallel transport, which is based on the local notion. 

Of course, it is possible to compromise: instead of choosing between sequence of elements "next to each other" or no sequence at all, we can construct a sequence of elements within a distance $\delta^2$ of each other. On the one hand $\delta^2$, as small as it is, still contains a lot of elements and thus it is statistically reliable. On the other hand, this resembles parallel transport. This, however, unnecesserely complicates the definition.  At the same time, for the purposes of the paper, it is sufficient if "parallel" elements are only parallel up to $\delta$ rather than $\delta^2$. For these reasons I chose to  abandon the idea of the sequence of elements (or parallel transport) and stick to quasilocal picture described earlier.

Now, lets go back to the definition of coordinate system. In light of the fact that each PST element corresponds to the \emph{timelike} vector in OST, all of the coordinate axes are timelike. Thus, a point in OST defined by these coordinates can be denoted as $(t_1, . . . , t_d)$. After having done that, the discretized version of linear combination can always be used to convert timelike coordinates to spacelike ones. Now a point in OST corresponds to a subset of PST (or, in other words, subset of a causal set $S$). That subset, of course, will depend on the choice of the origin (which corresponds to the OST position of the element $p_0$) as well as coordinate axes, which corresponds to the elements $p_1, . . . , p_d$.

Finally, in order to take into account the dependence on the small OST parameter $\delta^2$ and large PST/OST parameter $\Lambda$, we put both as a lower indices of $F$. The parameter $\delta^2$ was introduced in order to accomodate the fluctuations due to discreteness. In light of the curvature, everything is constrained to the neighborhood of the size that is first order small (in particular, $t_k$ are first order small, as well). Thats why the allowed error in their measurement is second order small, which is why this is denoted by $\delta^2$ rather than $\delta$. In order to save letters in the alphabet, $\delta$ will continue to be used to denote the first order small size of the local neighborhood. By making $\delta$ sufficiently small, one can make sure that $\delta^2$ satisfies expected properties of second-order-small parameter, without a need of multiplying it by any coefficient.  Finally, in order to define parallelism, we need to use the definition of non local relative velocity, provided in section 2.6. That definition was relying on an integer constant $n$. This gives us a final notation for an OST point: $F_{\Lambda, \delta^2, n} (p_0 ; p_1, . . . , p_d; t_1, . . . , t_d)$. As stated earlier, a point in OST is a subset in PST. In light of the fact that $S$ is PST, this is a subset (\emph{not} an element) of $S$. 

Now, in light of the curvature, the OST translations do not commute. Therefore, in order to define $F_{\Lambda, \delta^2, n} (p_0; p_1, . . . , p_d; t_1, . . . , t_d)$ we have to specify in what order the above mentioned translations are made. Roughly speaking, in order to arrive at any element of the set $F_{\Lambda, \delta^2, n} (p_0; p_1, . . . , p_d ; t_1, . . . t_d)$, one has to start from the location of $p_1$ and follow the direction of $p_0$ for the time duration $t_1$; then one has to follow the direction of $p_2$ for the time duration $t_2$ and keep going in this fashion, until, in the last step, one has to follow the direction of $p_n$ for the duration $t_n$. The words "follow a direction of $p_k$" imply a future geodesic translation by using some element "parallel to" $p_k$, where the quasi-local notion of parallelism, discussed earlier, is used. A couple of pages later we will postulate approximate transitivity, which is roughly
\beq (t_1, . . ., t_d) + (t_1', . . ., t_d') \approx (t_1 + t_1', . . . , t_d + t_d') \eeq
and show that it implies approximate commutativity, that is,
\beq F_{\Lambda, \delta^2, n} (p_0; . . . , p_i , . . . , p_j, . . . ;  . . . , t_i , . . . , t_j, . . . ) \approx F_{\Lambda, \delta^2, n} (p_0; . . . , p_j , . . . , p_i, . . . ;  . . . , t_j , . . . , t_i, . . . ) \eeq
where approximation is defined in appropriate sense. This means that arbitrary choice of order in which direction to go first and which to go next has only negligeable impact on the parametrization that we obtain.

Now, lets look a little bit more closely at the sequence of steps we just described. Suppose we first made a step in a direction $p_1$ and then $p_2$. In this case, at the very end of the first step and at the very beginning of the second step, we are at exact same location according to OST, but not according to PST. Thus, from PST point of view, we really made three or four steps. First we started at $q_1$. Then we moved to $q_2'$, which is a future geodesic translation of $q_1$. Since $q_2'$ is parallel to $p_1$, in order for our next step to be parallel to $p_2$ we need to rotate to $q_2$ which is parallel to the latter. Then we make a step in the direction of $q_2$. We could stop here, in which case we have made three steps. Or, if we want, we can perform an additional rotation after that step since the definition of $F_(\Lambda, \delta^2, n) (p_1, p_2; t_1, t_2)$ allows it. Likewise, at the very beginning, before making the step from $q_1$ to $q_2$, one could have possibly made a step from $q_1'$ to $q_1$, which turins it into total of $5$ terms. In any case, we had no choice as far as the rotation between first and second step is concerned. Thus, we had to make at least three and at most five steps.  

Similarly, the $n$-step OST sequence of points will become $2n$-step sequence from PST point of view. In this sequence, $q_k$ is a geodesic future translation of $q_{k-1}'$, and $q_k'$ is a rotation of $q_k$. Now, if $q_k$ and $q_k'$ move too close to the speed of light with respect to each other's reference frame, then, as a result of relativistic effects, the small fluctuations in OST-projection of Poisson scattering will result in very large OST fluctuations of $q_k$ in the reference frame of $q_k'$ and vise versa. In order to avoid this, their relative velocities are bounded by $\Lambda$. It should be understood, though, that $\Lambda$ can be very large, possibly close to the speed of light. This brings us to the following definition: 

\textbf{Definition}: Let $p_0; p_1, . . . , p_d$ be elements of causal set and let $t_1, . . . , t_d$, as well as $\Lambda$ and $\delta$  be real numbers. Then $F_{\Lambda, \delta, n} (p_1, . . . , p_d ; t_1, . . . t_d) \subset S$ is a subset of $S$ defined as follows: $s$ is an element of $F_{\Lambda , \delta, n} (p_0; p_1, . . . , p_d ; t_1, . . . t_d)$ if and only if there exist $q_1, . . . , q_{d+1}$ and $q'_1, . . . , q'_{d+1}$ satisfying the following conditions:

a) $v_{p_i} (q_i) < \delta$

b) $max (\vert r_{q_i} (q'_i) \vert , \vert t_{q_i} (q'_i) \vert) < \delta^2$

c) $\vert v_{q'_i, n} (q_i) \vert < \Lambda$

d) $q'_{i+1} \in G_{t_i} (q_i)$

e) $q_0'=p_0$ and $q'_{d+1} =s$

Note that in the above definition the velocity is bounded by $\delta$ while the position by $\delta^2$. That is due to the fact that the local neighborhood constraints velocity to $\Lambda$ and position to $\delta$. Thus, as far as velocity is concerned, $\delta$ is small compared to $\Lambda$; and, as far as position is concerned, $\delta^2$ is small compared to $\delta$. 

Since each of the PST elements has both well defined position and well defined velocity, we can get rid of element $p_0$ by allowing $p_1$ be used for both $p_0$ and $p_1$. More specifically, its OST projection will serve the purpose of $p_0$ while its PST/OST projection will serve a purpose of $p_1$:
\beq F_{\Lambda , \delta, n} (p_1, . . . , p_d ; t_1, . . . t_d) = F_{\Lambda , \delta, n} (p_1; p_1, . . . , p_d ; t_1, . . . t_d)  \eeq
Of course if we do that, then even in the flat scenario, $F_{\Lambda, \delta^2, n} (p_{a_1}, . . . , p_{a_d}; u_{a_1}, . . . , u_{a_d})$ depends on the choice of $a_1$, although it is true that, for any given $a_1$ it is independent of the choice of $a_2, . . . , a_d$. What comes to our rescue is that, in light of the curvature, we have already allowed the error $\epsilon \delta$ in our definition. As a side benefit to that, if we make sure that these elements are separated by the distance much smaller than $\epsilon \delta$, they \emph{would} commute. 

We would now like to define what it means for that choice of PST elements to span the subset $T \subset S$. This will be defined as follows: 

\textbf{Definition:} Let $\{p_1, . . . , p_d \}$ and $T$ be subsets of $S$, let $\Lambda$, $\delta$, $\epsilon$ and $\sigma$ be real nubmers and let $n$ be an integer. We say that $p_1, . . . , p_d$ \emph{span} $T$ up to $(\Lambda, \delta, \epsilon, \sigma, n)$ if the following is true: 

a) For any element $q \in T$, there is at least one choice of $t_1, . . . , t_d$, satisfying $max (t_1, . . . , t_d) < \delta$ such that $q \in F_{\Lambda, \delta^2, n} (p_1, . . . , p_d; t_1, . . . , t_d)$ 

b) Let $U$ be defined as follows: $q$ is an element of $U$ if and only if there is at least one element $q'$, and at least one choice of $t_1, . . . , t_d$, $t_1', . . . , t_d'$ and $t_1'', . . . , t_d''$ such that 

(i) $q \in F_{\Lambda, \delta, n} (p_0; p_1, . . . , p_d; t_1, . . . , t_d)$

(ii) $q' \in F_{\Lambda, \delta, n} (p_0; p_1, . . . , p_d; t_1', . . . , t_d')$

(iii) $q \in F_{\Lambda, \delta, n} (q'; p_1, . . . , p_d; t_1'', . . . , t_d'')$

(iv) There is at leat one $k$ such that $\vert t_k -t_k'-t_k'' \vert > \epsilon \delta$

Then $\sharp U < \sigma \sharp T$. 

The purpose of part a is to say that the choice of basis is complete. We note that in the above definition no accommodation for "missing" some elements of $T$ due to discrete fluctuations was made. That is because the definition of $F_{\delta^2, \Lambda} (p_0; p_1, . . . , p_d; t_1, . . . , t_d)$ incorporated the "width" $\delta^2$ of the paths involved. Thus, while the paths of one-element width might miss a lot of elements, the paths of width $\delta^2$ would not (in light of the constant $\tau_0$ in the definition of distances, one-element width is the same thing as the width $\tau_0$ and it is assumed that $\tau_0 << \delta^2$, which means that width $\delta^2$ corresponds to large number of elements). In fact, saying that a path of one element width would miss a given element but still hit its vicinity is equivalent to saying that a path of larger width would hit that element itself. That is why only one of the two provisions is necessary, but not both. 

As far as part b is concerned, $\sigma$ is assumed to be a small constant. Thus, $U$ consists of rare elements that are only there due to bizzare discrete fluctuations, and in continuum case won't exist at all. Now, (i), (ii) and (iii) are just the definition of notation, and they are summed up in the assumption
\beq (t_1, . . ., t_d) \approx (t_1', . . . , t_d') + (t_1'', . . . , t_d'') \eeq
which, of course, we are allowed to make. So the key criteria for $U$ is (iv), which says that 
\beq (t_1', . . . , t_d') + (t_1'' . . . t_d'') \not\approx (t_1'+t_1'', . . ., t_d' + t_d'') \eeq
where the criteria for $\approx$ is $\epsilon \delta$. After all, since we are restricting ourselves to small $\delta$ neighborhood (in order to avoid curvature), any meaningful approximation has to be an approximation up to something second order small, such as $\epsilon \delta$. Thus, when we say that $U$ contains very few elements, we are implying that in most cases  
\beq (t_1', . . . , t_d') + (t_1'' . . . t_d'') \approx (t_1'+t_1'', . . ., t_d' + t_d'') \eeq
up to $\epsilon \delta$. This, in particular, implies commutativity. For example, if $d=2$, then 
\beq F_{\Lambda, \delta^2, n} (r_1, r_2; t_1, t_2)= (t_1, 0) + (0, t_2) \eeq
and
\beq F_{\Lambda, \delta^2, n} (r_2, r_1; t_2, t_1)= (0, t_2) + (t_1, 0) \eeq
Then, \emph{if we apply transitivity to both equations} we get 
\beq F_{\Lambda, \delta^2, n} (r_1, r_2; t_1, t_2)= (t_1+0, 0+t_2) \eeq
and 
\beq F_{\Lambda, \delta^2, n} (r_2, r_1; t_2, t_1)= (0+t_1, t_2+0) \eeq
The lat two equation, of course, imply the desired result
\beq F_{\Lambda, \delta^2, n} (r_1, r_2; t_1, t_2)=F_{\Lambda, \delta^2, n} (r_2, r_1; t_2, t_1) \eeq
Thus, in the above derivation the only non-trivial step was the tranitivity. Thus, the fact that we have postulated transitivity as part of our criteria assures commutativity which in turn. As mentioned earlier, the latter is a key ingredient of the meaningfulness of the definition of coordinate system -- after all, if it wasn't for commutativity there would be no justification for the particular choice I have made on the sequence of transitions.  

We have just defined what it means for the set of elements to span a certain region. Now, do we intend to postulate that \emph{most} choices of $d$ elements span a region? Of course not! In fact, the vast majorit of such choices are separated arbitrary far from each other, which means that the curvature does not allow this to happen. We do want to claim, however, that most \emph{local} selections of $d$ elements span a region. We formally define the set of all possible \emph{local} choices of $d$ elements of $S$ by \emph{branching} of $S$, as follows: 

\textbf{Definition}: Let $S$ be a causal set and let $\Lambda$ and $\delta$ be real numbers and let $n$ and $d$ be an integer. Then \emph{$(d; \Lambda, \delta, n)$-branching} of $S$, which is denoted by $B_{d; \Lambda, \delta, n} (S)$ is a subset of $S^d$, denoted by $B_{d; \Lambda, \delta, n} (S)$, and is defined to be a set of all $d$-tuples of elements $(p_1, . . ., p_d)$ satisfying
\beq max (\vert t_{p_i} (p_j) \vert , r_{p_i}(p_j) ) < \delta \eeq
and
\beq max ( \vert v_{p_i; n} (p_j) \vert ) < \Lambda \eeq

Now we would like to define \emph{what} region are we wanting the branching to span. We would define that local region to be a \emph{cylinder} around one of these elements of a specified size, where a cylinder, whose axis coincides with the direction of OST vector corresponding to that particular PST element. This cylinder is defined as follows:

\textbf{Definition}: Let $S$ be a causal set and let $h$, $r$ and $\Lambda$ be integers. Then, for any $p \in S$, the \emph{cylinder} $C_{h, r; \Lambda} (p)$ consists of points $q$ for which $\vert t_p (q) \vert < h$, $\vert r_p (q) \vert < r$ and $\vert v_p (q) \vert < \Lambda$. 

In order to save letters, we will assume $h= r = \delta$ from now on. We now will define the set of $d$-tuples that map all of the $d$ cylinders defined by each of its elements to be \emph{map generators}:

\textbf{Definition} Let $p_1, . . . , p_d$ be elements of $S$, let $\Lambda$, $\delta$, and $\sigma$ be real nubmers and let $n$ be an integer. We say that $\{p_1, . . . , p_d\}$ is a \emph{map generator} of $S$ up to $(\Lambda, \delta, \sigma, n)$ if for every $k$, $\{p_1, . . . , p_d\}$ spans $C_{\delta, \delta; \Lambda} (p)$ up to $(\Lambda, \delta, \epsilon, \sigma, n)$. The set of all map generators up to $(\Lambda, \delta, \sigma, n)$ will be denoted by $m (d; \Lambda, \delta, \sigma, n)\subset S^d$ 

Finally, we are ready to define what we mean by $S$ being $d$-dimensional:

\textbf{Definition} Let $\Lambda$, $\delta$, and $\sigma$ be real nubmers and let $d$ and $n$ be an integers. $S$ is said to be \emph{d-dimensional} up to $(\Lambda, \delta, \sigma, n)$ if 
\beq \frac{\sharp (B_{d; \Lambda, \delta, n} (S) \setminus m_{d; \Lambda, \delta, \sigma, n})}{\sharp B_{d; \Lambda, \delta, n} (S)} < \sigma \eeq

It is important to note that in standard causal set literature the notion of dimension is defined in a much looser sense and it relies on various statistical properties which are well defined for all causal sets: both manifoldlike and non-manifoldlike ones. However, for the purposes of this paper, that is useless as far as defining the notion of PST likeness is concerned. Thus, much stricter criteria for a dimensionality is introduced, which singles out a very small class of causal sets; the PST-like causal sets will be defined as even smaller subclass of that class. 

\noindent{\bf 3.5 Distance properties of PST extension of 1+1 dimensional submanifold of OST}

Despite the fact that we have found a way to make sure that causal set has local $d$-dimensional OST coordinate system, this does not yet prove that it is either OST-like or PST-like from metric perspective. In fact, we are yet to see whether simple things like Pythagorean theorem or Lorentzian geometry hold. As stated earlier, we would like to avoid actually postulating Pythagorean theorem or any other ingredients of the notion of a metric, and instead postulate some geometrical notions that are more intuitive to grasp, and select them in such a way that they ultimately imply the sought-after properties of a metric.

We know that if we consider scattering of points in $1+1$ dimensional OST, we will get the expected properties of the Lorentzian distance. In the PST case, in light of the fact that the fictitious photon moves with near-lightlike velocity within a very short time interval of its imaginary emission, the result of OST-based causal set theory is carried over on a larger scale. This suggests that we can employ some constructions in $d$ dimensional OST that involves $1+1$ dimensional OST planes (each of which will then be extended to PST through tangent vectors) in order to use the $1+1$ dimensional geometry as a stepping stone to obtain the $d$ dimensional one (here, as well as throughout remainder of this paper, when we speak of certain number of dimensions, we are referring to OST ones, unless otherwise specified). 

Now, in $1+1$ dimensions it might still not be as easy as it seems. In order to be able to talk about $1+1$ dimensional Poisson scattering, we have to know what we mean by "uniform density". In order to say that density at $(t_1, x_1)$ is the same as density at $(t_2, x_2)$, we have to say that a square defined by 
\beq \{(t_1- \delta t/2, x_1 - \delta x/2 ),(t_1- \delta t/2, x_1 + \delta x/2 ),(t_1+ \delta t/2, x_1 - \delta x/2 ),(t_1+ \delta t/2, x_1 + \delta x/2 ) \} \eeq
 gets as many points as the square defined by 
\beq \{(t_2- \delta t/2, x_2 - \delta x/2 ),(t_2- \delta t/2, x_2 + \delta x/2 ),(t_2+ \delta t/2, x_2 - \delta x/2 ),(t_2+ \delta t/2, x_2 + \delta x/2 ) \} \eeq
It is very important that these squares are much smaller than the original $\delta$-size neighborhood we are concerned about. After all, if we wanted to, we could have replaced $t$ and $x$ by $sin \; t$ and $sin \; x$ in the above definitions; this would, clearly, imply non-uniform density. Now, if we were to look at the number of elements received at the square of size $\delta$, then there would be no way to distinguish the two. But, by partitioning $\delta$ into the smaller intervals of size $\delta^2$ we can, in fact, distinguish these two cases: in one case we have 
\beq t= m \delta^2 \; , \; x = n \delta^2 \eeq
 and in another we have 
\beq sin \; t= m \delta^2 \; , \; sin \; x = n \delta^2 \eeq
 as defining features of the size of small square. Now since we have postulated transitivity in the last section, we can focus on a single square around local origin,
\beq \{(- \delta t/2,  - \delta x/2 ),(- \delta t/2, \delta x/2 ),(\delta t/2, - \delta x/2 ),(\delta t/2,  \delta x/2 ) \} \eeq
and the rest will follow by transitivity. 

To transform the above considerations into rigourous definition, we have to define a class of two dimensional subsets of $S$ (or, in other words, plane patches embedded into higher dimensional spacetime) which, on average, suppose to receive equal number of elements. We proceed in the following way: first, as expected, we select two PST elements (or, equivalently, two OST vectors) $p$ and $q$ to define two-dimensional coordinates of the plane in which the parallelogram sits. If we were interested in the OST topology, rather than the PST one, we could build the parallelogram out of "thickened points", where by a "thickened point" we mean a region of the size $\delta^3$. Thus, the parallelogram would look like this:  
\beq U_{\Lambda, \delta^2, n} (p, q) = \bigcup_{\vert t_1 \vert < \delta^2 \; ; \; \vert t_2 \vert < \delta^2} F_{\Lambda, \delta^3, n}(p, q; t_1', t_2') \eeq
Here, the size of entire region is $\delta$ in order to avoid curvature, the size of small sample square is $\delta^2$ in order to distinguish $t$ and $x$ from $sin \; t$ and $sin \; x$ (see discussion above), and the size of thickened point is $\delta^3$ in order for it to measure the structure of the square of size $\delta^2$ precisely enough (after all if thickened point had a size $\delta^2$ we would have an error of the order $\delta^2$ in measuring the size of the square, and thus we could mistake a square of size $\delta^2$ with a square of size $2 \delta^2$ and get the density off by a factor of $2$).  

Now, since we are interested in PST topology, we have to add similar restriction on velocities. On the one hand, the smallness of the local neighborhood implies that the velocities have to be restricted to the range of a very small width, $\delta$. On the other hand, similar to space displacement mentioned in the previous paragraph, that range can be displaced from velocities of $p$ and $q$ by very large amounts, as long as that displacement doesn't exceed some large number $\Lambda$ (as before, $\Lambda$ is assumed to be very close to speed of light since there are no curvature effects in velocity direction; at the same time, $\Lambda$ is assumed to be slightly less than $c$ in order to avoid the situation when the spacing between points in Poisson distribution appears to be large due to relativistic effects). Thus, we need some third vector, $s$ to define the velocity range
\beq V_n (s; \delta) = \{ e \vert \vert v_{s;n} (e) \vert < \delta \} \eeq
Therefore, the generic parallelogram becomes
\beq W_{\Lambda, \delta, n} (p, q, s; t_1, t_2, \delta) = U_{ \Lambda, \delta^2, n}(p, q; t_1, t_2, \delta) \cap V_n (s; \delta) \eeq
where it is understood that 
\beq max (\vert t_p (s) \vert, \vert r_p (s) \vert, \vert t_q (s) \vert, \vert r_q (s) \vert )< \delta \nonumber \eeq
\beq max \{ \vert t_p q \vert, \vert r_p q \vert \} < \delta \epsilon \eeq
Then, in order to obtain 1+1 dimensional geometry, we will go ahead and define the average of $1+1$-dimensional "densities" of Poisson distribution over all of these 1+1 dimensional parallelograms.  This, of course, would not be possible in the continuum scenario, but it can be accomplished in discrete case:   
\beq \rho_{\Lambda, n} (t_1, t_2, \delta) = \frac{\sum_{(p, q) \in B_{2; \Lambda, \delta, n} (S)} \sharp F_{\Lambda, \delta^2, n} (p, q; t_1, t_2)}{\sharp B_{2; \Lambda, \delta, n} (S)} \eeq
Finally, we impose a constraint that the global average of the density, as defined above, is restricted to some, possibly narrow, range of numbers. This, together with transitivity, implies a uniform probability density on each small 1+1 dimensional region. At the same time, the fact that the constraint is only imposed on global average of the density, allow for large local fluctuations, just as normally expected of Poisson distribution.  

This, however, is not enough to obtain relativity on a timelike two dimensional surface. The other piece of information we need is a constant speed of light on that surface. Since both of our coordinates are timelike, in order to define speed of light we use the Lorentz transformations 
\beq t = \gamma_1 t_1 + \gamma_2 t_2 \; ; \; x = \gamma_1 v_1 t_1 + \gamma_2 v_2 t_2 \eeq
Now, since the relation between $\gamma_i$ and $v_i$ is derived by means of Pythagorean theorem, which we are not supposed to know, we are forced to replace $\gamma_1$, $\gamma_2$, $\gamma_1 v_1$ and $\gamma_2 v_2$ with $k_1$, $k_2$, $k_3$ and $k_4$ respectively, where the latter are four \emph{independent} numbers.

Furthermore, due to the random fluctuations, two things will happen. First of all, the lightcone is not exact but only approximate. And secondly, the approximate light cone is not true all the time but only most of the time. Thus, we define a "linear portion" to be a set of pairs of points for which the approximation holds, within a specified degree of precision, and then we simply say that out of the sample of all pairs of elements that are close to each other, the percentage of the ones that don't belong to "linear portion" is very small. This, formally translates into two definitions: 

\textbf{Definition}: Let $\delta$ and $\epsilon$ be real numbers and $n$ be an integer. A \emph{linear portion} of $S$ based on $(\Lambda, \delta, \epsilon, n)$ is a subset of $S \times S$, denoted by $L_{\Lambda, \delta, \epsilon, n} (S)$ which consists of all pairs $(p, q)$ for which one can find constants $k_1, . . . , k_4$ such that, whenever $a$ and $b$ satisfy 
\beq max ( \vert r_p (a) \vert , \vert t_p (a) \vert , \vert r_q (a) \vert , \vert t_q (a) \vert , \vert r_p (b) \vert , \vert t_p (b) \vert , \vert r_q (b) \vert , \vert t_q (b) \vert ) < \delta \eeq
and
\beq min ( \vert v_{p; n} (a) \vert, \vert v_{q; n} (a) \vert ) < \Lambda \eeq
the following statements are true:

a) If $a \prec b$, then $t_p (a) < t_p (b)$ and $t_q (a) < t_q (b)$. Likewise, if $b \prec a$, then $t_p (b) < t_p (a)$ and $t_q (b) < t_q (a)$ 

b) If either $a \prec b$ or $b \prec a$, then 
\beq \vert k_1 (t_p (b) - t_p (a)) + k_2 (t_q (b) - t_q (a)) \vert > \vert k_3 (t_p (b) - t_p (a)) + k_4 (t_q (b) - t_q (a)) \vert - \epsilon \delta \eeq

c) If $a$ and $b$ are causally un-related, 
\beq \vert k_1 (t_p (b) - t_p (a)) + k_2 (t_q (b) - t_q (a)) \vert < \vert k_3 (t_p (b) - t_p (a)) + k_4 (t_q (b) - t_q (a)) \vert + \epsilon \delta \eeq

\textbf{Definition}: Let $S$ be causal, let $\Lambda$, $\delta$, $\epsilon$ and $\rho$ be real numbers, let $n$ be an integer, and let $L_{\Lambda, \delta, \epsilon} (S)$ be linear portion of $S$ based on $(\Lambda, \delta, \epsilon, n)$. Then the set $S$ is said to be \emph{linear} up to $(\Lambda, \delta, \epsilon, \sigma)$ if 
\beq \frac{ \sharp (B_{2; \Lambda, \delta, n} (S) \setminus L_{\Lambda, \delta, \epsilon, n} (S)) }{ \sharp B_{2; \Lambda,\delta, n} (S)} < \sigma \eeq

Now lets go back to the coefficients $k_1, . . . , k_4$. As mentioned earlier, these coefficients correspond to $\gamma_1$, $\gamma_2$, $\gamma_1 v_1$ and $\gamma_2 v_2$. The reason they are viewed as completely independent of each other is that we are not allowed to use Pythagorean theorem, which means we don't have means of doing the usual derivations that are used to prove the dependence of $\gamma$ on $v$. However, there is a geometric way of making this independence compatible with what one would expect of scattering when Pythagorean theorem \emph{does} hold. Namely, the dependence of $\gamma$ on $v$ only holds if we assume $g_{\mu \nu} v^{\mu} v^{\nu} = 1$. Thus, the independence of the four coefficients is equivalent to abandoning that assumption. This means that we can, statistically, enforce the expected dependence of $\gamma$ on $v$ by "constraining" the "linear density" of our paths to be similar. This can be done with the following definition

\textbf{Definition}: Let $\rho$, $\sigma$, $\delta$ and $\epsilon$ be real numbers. A causal set $S$ is \emph{linearly uniform} up to $(\sigma, \delta, \epsilon)$ with \emph{linear density} $\rho$ if, for every $m< \delta$ (\emph{both positive and negative}), 
\beq \frac{ \sharp \{ (p, q) \vert q \in G_m (p) \wedge ( \tilde{\tau} (p, q) < (\rho - \epsilon \delta) \vert t_p (q) \vert \vee  \tilde{\tau} (p, q) > (\rho + \epsilon \delta) \vert t_p (q) \vert ) \} }{\sharp \{ (p, q) \vert q \in G_m (p) \} } < \sigma \eeq
where $\tau (p, q)$ is the largest possible $n$ for which one can find a sequence $p \prec r_1 \prec . . . \prec r_{n-1} \prec q$ and $\tilde{\tau} (p, q) = max (\tau (p, q), \tau (q, p))$.

Up to this point we were rightly assuming that distance is defined by longest path. But we ignored the fact that some of the paths that connect two PST elements on a plane might not lie on that plane. In order to fix that, we have to postulate that as long as the two elements lie on an OST $1+1$ dimensional plane, the longest possible path that connects them lies on that $1+1$ dimensional plane as well. Equivalently, this means that the longest path between two timelike separated elements lies on the intersection of all possible OST planes that pass through these elements. We will do this by first defining a notion of a pair of elements for which it is true in its neighborhood, and then postulate that most (though not all) of the elements of $S$ are such elements. Thus, it boils down to two definitions: 

\textbf{Definition}: Let $p$ and $q$ be elements of $S$, let $n$ be an integer and let $\Lambda$, $\delta$, and $\epsilon$ be reals. Then $(p, q)$ is said to be \emph{shortcut free} up to $(\Lambda, \delta, \epsilon, n)$ if whenever there is a sequence of elements $r_1 \prec . . . \prec r_m$ such that $r_1$ and $r_m$ are both elements of $F_{\delta , \Lambda, n} (p, q; t_1, t_2)$, there is also a sequence of elements $s_1 \prec . . . \prec s_{m'}$ such that $r_1 = s_1$, $r_m = s_{m'}$, $s_k \in F_{\Lambda, (\epsilon \delta), n} (p,q)$ and $\tau_0 m' > \tau_0 m - \epsilon \delta$, where $\tau_0$ is an atomic scale mentioned in 2.1 (it was introduced in order to say $\delta^2 << \delta$).

\textbf{Definition}: Let $\Lambda$, $\delta$, $\epsilon$ and $\rho$ be reals, and let $n$ be an integer. Let $A_{\Lambda , \delta, \epsilon, n} (S) $ be a subset of $B_{2; \Lambda, \delta, n}$ consisting of pairs of elements that are shortcut-free up to $(\Lambda, \delta, \epsilon)$ Then causal set $S$ is said to be \emph{shortcut free} up to $(\Lambda, \delta, \epsilon, \sigma)$ if
\beq \frac{\sharp(B_{2; \Lambda, \delta, n} (S) \setminus A_{\Lambda, \delta, n} (S))}{\sharp B_{2; \Lambda, \delta, n} (S)} < \sigma \eeq

\noindent{\bf 3.6 Distances in d OST dimensions}

We are now going to try and see whether we can go from two dimensions to $d$ dimensions based on just what we have; or, if not, what additional axioms should we add. In order to do that, we are going to make some loose jumps and make a geometrical argument that we would have made if we knew we were dealing with usual continuum geometry, and then we will go back and postulate just enough of the stuff that we need to be able in order to make the argument we just heard ourselves making.

We have already established that, up to some statistical fluctuations, if the PST-based causal set $S$ satisfies all of the above definitions then there is a local (in terms of OST) choice of $d$ elements of PST which can be viewed as timelike vectors in OST that form a basis of a coordinate system. Lets call them $v_0^{\mu}, . . ., v_{d-1}^{\mu}$. They can be thought of as timelike, non-orthogonal, version of vierbines. 

Now, we are going to do a sort of Gramm Schmidt process. We start with out original element of interest,
\beq w_{d-1}^{\mu} = \sum_{\nu =0}^{d-1} a_{d-1; \nu} v_{\nu} ^{\mu} \eeq
Here, Einstein's summation convention was not used because $a_{\rho \sigma}$ are just coefficients, and they are not part of the vector or tensor. Now, we let $w_{d-2}^{\mu}$ be a projection of $w_{d-1}^{\mu}$ onto the $d-1$ dimensional hyperplane spanned by $v_0^{\mu}$ through $v_{d-2}^{\mu}$. Then we can view the space spanned by $w_{d-1}^{\mu}$ and $w_{d-2}^{\mu}$ as a two dimensional plane. We then use the statistical argument on a plane to show that the distances work as expected, and, in particular, $w_{d-1}^{\mu} w_{d-1; \mu} = w_{d-2}^{\mu} w_{d-2; \mu} - e_{d-1}^{\mu} e_{d-1; \mu}$, where $e_{d-1}^{\mu} = w_{d-1}^{\mu} - w_{d-2}^{\mu}$. Then, we similarly define $w_{d-3}^{\mu}$ to be a projection of $w_{d-2}^{\mu}$ on the space spanned by $v_0^{\mu}$ through $v_{d-3}^{\mu}$ and repeat similar argument, and keep going until we reach $0$. By induction, it is clear that we will get the desired metric.

Now, lets go back and look at some of the gaps in the above argument that we are going to fill with axioms. Since we have already established that on the plane the geometry works, the only time we made a lose gap was \emph{before} we knew we were dealing with a plane. One assumption that was made was that there is a unique PST element (or, equivalently, OST vector) $w_{d-1}$ corresponding to the linear combination of different elements $v_k$. Now, as was extensively discussed earlier, it is true that one can travel specified time intervals $a_{d-1; k}$ in the directions parallel to $v_k$ (in terms of non local relative velocities of section 2.6 \emph{as opposed to} parallel transport), in specified order. However, while this would get one to a specific location, this, technically, has nothing to do with producing PST element $w_{d-1}^{\mu}$ since the latter is a vector at a point in OST, not a segment between two points. Thus, we have to assume some kind of one to one correspondence between segments and vectors that point in the direction given by that segment. This is accomplished by imposing the following restriction:

\textbf{Definition}: Let $S$ be a causal set and let $\Lambda$, $\delta$ and $\epsilon$ be reals and let $n$ be an integer. $S$ is said to be \emph{exponential} up to $(\Lambda, \delta, \epsilon, n)$ if for any $p \in S$ and $q \in S$ satisfying
\beq max (\vert r_p (q) \vert, \vert t_p (q) \vert, \vert r_q (p) \vert, \vert t_q (p) \vert) < \delta \eeq
and
\beq max ( \vert v_{p; n} (q) \vert, \vert v_{q; n} (p) \vert ) < \Lambda \eeq
there exist $p'$ and $q'$ such that both $p' \in G(q')$ and $q' \in G (p')$ holds, which satisfy
\beq max (\vert r_p (p') \vert, \vert t_p (p') \vert, \vert r_{p'} (p) \vert, \vert t_{p'} (p) \vert) < \delta \epsilon \eeq
and
\beq max (\vert r_q (q') \vert, \vert t_q (q') \vert, \vert r_{q'} (q) \vert, \vert t_{q'} (q) \vert) < \delta \epsilon \eeq

In the above definition, $p'$ and $q'$ are rotations of $p$ and $q$, respectively. So, what it says, is that there is a way to "rotate" both OST vectors to "line up" with the geodesics that connects them. While in smooth geometry, $p' \in G(q')$ and $q' \in G(p')$ are equivalent statements, in discrete case, as mentioned at the end of section 2.4, such is not the case. That is why in the above definition we had to explicitly say that \emph{both} of them hold. 

Now, after we obtained $w_{d-1}^{\mu}$ another assumption was made. Namely, it was assumed that one can take a projection of $w_{d-1}^{\mu}$ onto a $d-1$ dimensional hyperplane spanned by $v_0^{\mu}$ through $v_{d-2}^{\mu}$ and also that that projection is unique. Furthermore, we are using the "perpendicular" direction of the projection operation as one of the coordinate axes; this, of course is a stronger statement than uniqueness, so once we take care of that, we will get uniqueness automatically. So we need two definitions. The first one is a definition of projection, and the second one is a definition of a certain property of projection that assures that it can be used as one of the axes. Now, in light of discreteness we don't really want to say that projection is unique. Rather, we would like to say that if two different elements are both good candidates for projection, then they are close to each other. This means that projection should be defined as a \emph{subset} (not an element) of a set $T$ we are doing the projection onto. This subset conists of elemnts that are almost coinciding with each other, and each of these elements approximates what we mean by projection. It is officially defined as follows: 

\textbf{Definition} Let $T$ be subset of $S$, $p$ be an element of $S$, let $\Lambda$, $\epsilon$, $\delta$, $\chi$ and $\lambda$ be reals, and let $n$ be an integer. A \emph{projection} of $p$ onto $T$, which is denoted by $P_{p;n,\Lambda  \delta^2, \epsilon^2, \chi^2, \lambda^2} (T) \subset T$ is a subset of $T$ consisting of elements $q$ such that 

a) $q \in T$

b) If $q' \in T$, $v_{q; n} (q') < \Lambda$ and $max (\vert t_q (q') \vert, r_q (q'))< \delta^2$, then $r_{q'} (p) > r_q (p) + \epsilon^2$

c) If $max (\vert t_q (q') \vert, r_q (q'))> \chi^2$ and $v_{q; n} (q') < \Lambda$  then there is at least one $q''$ such that $max (\vert t_{q'} (q'') \vert, r_{q'} (q''))< \delta^3$ and $r_{q''} (p) < r_q (p) - \lambda^2$

d) $t_q (p) < \epsilon^2$

The above definition is taylored for a specific case where $T$ looks like a timelike hyperplane, even though it weren't explicitly stated. In part b of the above definition, $q'$ is a Lorentz rotation of $q$. If $q$ points in a direction that is parallel to the plane, while the segment connecting $p$ and $q$ is perpendicular, it is easy to see that due to Lorentz transformations we would expect $r_{q'} (p) > r_q (p)$ in a flat continuum caes. So, if we take into account the curvature as well as fluctuations due to discreteness, we instead get $r_{q'} (p) > r_q (p) + \epsilon^2$ which is a somewhat weaker statement. 

On the other hand, part c of above definition talks about a scenario where $q'$ is a displacement of $q$ as well, rather than pure rotation. In this case, due to translation the Lorentzian distance between $p$ and $q'$ is smaller. Now, this doesn't necessary say anything about the comparison of $r$ or $t$ coordinates since we don't know in what way $q'$ is rotated relative to $q$, if at all. However, we do know that if we explore all possible rotations of $q'$ we will be able to make $r$ smaller than if we were explored all possible rotations of $p$ and $q$. Now since in the latter case $q$, itself, is assumed to minimize $r$ (due to part a), thats why in part b I was comparing $q''$ to $q$ instead of $q''$ to, say, $q^*$.  

Now, in light of $\Lambda$ part b might not be true if $q'$ is rotated relative to $q$ by an angle much greater than $\Lambda$ since in this case we won't have enough range to rotate $q'$ to the desired vector $q''$. However, things do work if all the elements of $T$ have relative velocity with smaller than $\Lambda$ with respect to each other. Such will be the case for $F_{\Lambda, \delta^2, n} (p_0; p_1, . . . , p_k; t_1, . . . , t_k)$, which is the ultimate example we are interested in. 

We will now proceed and define the situation where projections can be trusted to form a coordinate system. We will use the same approach as before. We will first define a class of points for which such projections are unique in their respective cylindrical neighborhoods. After that, we will postulate that this set takes up most of $S$ (and some small parameter $\sigma$ will be used to define the word "most"). Now, we will denote a hypersurface spanned by $p_1, . . ., p_k$ to be
\beq F_{\Lambda, \delta, n} (p_0; p_1, . . . ,p_k) = \bigcup_{\forall i (\vert t_i \vert < \delta)} F_{\Lambda, \delta^2, n} (p_0; p_1, . . . ,p_k; t_1, . . . ,t_k) \eeq
Here, $k$ was used instead of $d$ because, as we repeatedly do Gramm Schmidt process, we will use all dimensions between $2$ and $d$; thus, $k (\leq d)$ will be used as a dimensional parameter in the definition. We would like to say that the direction "perpendicular" to $F_{\Lambda, \delta, n} (p_0; p_1, . . . ,p_k)$ \emph{as defined by projection}, "commutes" with any direction parallel to $F_{\Lambda, \delta, n} (p_0; p_1, . . . ,p_k)$. That will be defined by saying that if we have two elements outside of that hyperplane, and both of them have the same distance to their respective projections, and also if both of them are pointing in the same direction as their projections do, then the time shift, radius and velocity of these two elements relative to each other almost approximate to respective relative coordinates of their projections. More formally, it can be written as follows:

\textbf{Definition} Let $p_0$ be an element of $S$, let $\delta$, $\epsilon$, $\chi$, $\lambda$ and $\Lambda$ be real number and let $n$ be an integer. $p$ is said to be \emph{projection compatible} up to $(\Lambda, \delta, \epsilon, \chi, \lambda, n)$ if the following is true: Suppose $s_1$ and $s_2$ are elements of $S$ satisfying $v_{p_0; \Lambda, n} (s_1) < \Lambda$, $v_{p_0; \Lambda, n} (s_2) < \Lambda$, $t_{p_0} (s_1) < \sigma$, $t_{p_0} (s_2) < \delta$, $r_{p_0} (s_1) < \delta$, and $r_{p_0} (s_2) < \delta$. Furthermore, suppose the following conditions are met: 

(i) $s_1' \in P_{s_1;n,\Lambda  \delta^2, \epsilon^2, \chi^2, \lambda^2} (F_{\Lambda, \delta, n} (p_0; p_1, . . . ,p_k))$ 

(ii) $s_2' \in P_{s_2;n,\Lambda  \delta^2, \epsilon^2, \chi^2, \lambda^2} (F_{\Lambda, \delta, n} (p_0; p_1, . . . ,p_k)) $ 

(iii) $max (v_{s_1' ; n} (s_1), v_{s_2' ; n} (s_2))< \delta$

(iv) $\vert r_{s_2'} (s_2) -  r_{s_1'} (s_1) \vert < \delta^2$

Then the following will be true:

a)$\vert r_{s_1'} (s_2') - r_{s_1} (s_2) \vert < \epsilon^2$

b)$\vert t_{s_1'} (s_2') - t_{s_1} (s_2) \vert < \epsilon^2$

c)$\vert v_{s_1'} (s_2') - v_{s_1} (s_2) \vert < \epsilon$

\textbf{Definition}: Let $S$ be a causal set, let $\delta$, $\epsilon$, $\chi$, $\lambda$, $\Lambda$ and $\sigma$ be real number and let $n$ be an integer. Let $H_{\Lambda, \delta , \epsilon, \chi, \lambda, n}$ be a subset of $S$ consisting of elements that are \emph{projection-compatible} up to $(\Lambda, \delta , \epsilon, \chi, \lambda, n)$. Then $S$ is said to be \emph{projection compatible} up to $(\Lambda, \delta , \epsilon, \chi, \lambda, n, \sigma)$ if 
\beq \frac{ \sharp (S \setminus H_{\Lambda, \delta , \epsilon, \chi, \lambda, n})}{\sharp S} < \sigma \eeq

\noindent{\bf 4: Fields on a causal set}

Throughout this chapter, the distances are going to be scaled by a small parameter $\tau_0$, so that they no longer have to be integers and the smallest distance is much less than $1$. This is done in order to make sure that, for small distance scales $\delta$, we have $\delta^2 << \delta$ which would allow us to define approximations. 

\noindent{\bf 4.1 Key advantages of PST-based fields over OST-based ones}

We will now introduce the matter fields on a causal set. As was mentioned earlier, one of the main conceptual issues in doing this is the lightcone singularity. Due to the fact that we need to define a discretized verion of the "derivative" of a field in order to define Lagrangian density at point $p$, we need to look at nearby points of $p$. But, since the Lorentzian neighborhood is defined by a vicinity of light cone, the point $p$ has infinitely many scattered points within its $\epsilon$-neighborhood, and most of them have arbitrary large coordinate separation from $p$. Thus, if we use Lorentzian distance as the only clue in defining discretized derivative, that derivative would take into account the values of the field at all the "far away" points we just mentioned, which means that the result will have nothing to do with actual local value of a derivative of a smooth function, if such existed. I consider this to be the major conceptual issue in defining quantum field theory in a causal set background, and adressing this issue is the main motivation for the approach proposed in this paper.

Some attempts to define quantum field theory on a causal set were already made in \cite{Johnston1}, \cite{Johnston2}, \cite{Proceedings} and \cite{paper5}. As far as \cite{Johnston1} and \cite{Johnston2} goes, scalar field propagators were, in fact, successfully defined without any light cone singularities. This is due to the fact that a propogator is a function of two points, rather than just one. Suppose we are interested in computting the propagator between points $p$ and $q$, where $p \prec q$. The only part of the future half of the lightcone of $p$ that is relevent is the one that is in the past of $q$. This allows us to get rid of the infinite volume we were concerned about. 

Physically, this can be viewed as a consequence of the fact that the line connecting the "emission" and "absorption" points (or, equivalently, the line connecting sources and sinks) defines a "preferred frame". However, once more than one propagator is introduced, one immediately faces the situation that these two "preferred frames" can move arbitrary close to speed of light relative to each other. In fact, this phenomenon implies that even if we only restrict ourselves to propagators across very small distance, we can still have infinitely many propagators since there are infinitely many points in vicinity of light cone to choose from. This, of course, is just one of the symptoms that indicates that the light cone singularity is back in the picture.  

This is related to a reason why it was much more natural to avoid lightcone singularity in \cite{Johnston1} and \cite{Johnston2} than in \cite{Proceedings} and \cite{paper5}. Since the Lagrangian formalism hides in itself the key principles of entire quantum field theory as we know it, it also "hides" the possibility of arbitrary many propagators due to virtual particles. This can be seen by the fact that a one-element set is an intersection of all possible two-element sets containing that element. Thus, Lagrangian density, which is a function of a single point $p$, is really a function of intersection of all possible pairs of points, that cointain $p$; thus, it includes all possible propagators which have $p$ as either emission or absorbtion point. Since the actual Lagrangian was used in \cite{Proceedings} and \cite{paper5} \emph{as opposed to} \cite{Johnston1} and \cite{Johnston2}, the former approach faces the lightcone issues and not the latter. This, of course, comes with a price that \cite{Johnston1} and \cite{Johnston2} fails to represent the real life with arbitrary many virtual particles. While, of course, such scenario can be described within \cite{Johnston1} and \cite{Johnston2}, its absence is not ruled out, which makes the theory incomplete. 

Now let's look more specifically at what happened in \cite{Proceedings} and \cite{paper5}. In \cite{paper5}, a vector valued function $v^{\mu} (\phi, x)$ was defined to be a direction in which $\phi$ varies the least in the vicinity of $x$. It was noticed that in case $\phi$ is linear, such direction coincides with a direction of gradient of $\phi$. This lead to defining the kinetic term of the Lagrangian as  
\beq {\cal L} = ( v^{\mu} (\phi, x) \partial_{\mu} \phi )^2 \eeq
This, in fact, naturally restores locality without sacrificing relativity. However, this creates another problem: the definition of $v (\phi, x)$ is very non-linear, nor can it be seen as a perturbation to linearity, either. Thus, one can not use that Lagrangian to compute propagators or draw Feynman diagrams, since all of these are based on linearity. I belive this to be the main reason why one is limitted to numerical methods when it comes to actually solving some physical problem on a causal set background. 

On the other hand, in \cite{Proceedings} such issue does not exist if $\phi$ is assumed to be linear. For any given point $p$, and any unit timelike vector $v$, the Lagrangian density was defined as 
\beq {\cal L} = \frac{A_d}{\tau^2} (\phi (exp_p (\tau v/2)) - \phi (exp_p (-\tau v/2)))^2 - \nonumber \eeq
\beq - \frac{B_d}{\tau^{d+2}} \int_{\alpha (exp_p (-\tau v/2), exp_p (\tau v/2))} d^d q ( \phi (r) - \phi (p) )^2 \eeq
where $\tau$ is some small constant and $A_d$ and $B_d$ are dimension (d) -dependent coefficients (here, $exp_p (kv)$ is an element of a smooth manifold one can reach by travelling distance $k$ from point $p$ along the geodesic whose tangent at $p$ is a unit vector $v$). It can be easilly shown that, for the case that $\phi$ is linear, 
\beq \frac{(\phi (exp_p (\tau v/2)) - \phi (exp_p (-\tau v/2)))^2}{\tau^2} = (\partial_0 \phi)^2 \eeq
and
\beq \frac{1}{\tau^{d+2}} \int_{\alpha (exp_p (-\tau v/2), exp_p (\tau v/2))} d^d q ( \phi (r) - \phi(p) )^2 = C_d (\partial_0 \phi)^2 + D_d \sum_{k=1}^{d-1} (\partial_k \phi)^2 \eeq
Thus, the Lagrangian takes the form 
\beq {\cal L} = (A_d - B_d C_d) (\partial_0 \phi)^2 - B_d D_d \sum_{k=1}^{d-1} (\partial_k \phi)^2 \eeq
Now, $C_d$ and $D_d$ are dictated by geometry, while $A_d$ and $B_d$ are up to us to define. So if we define them as 
\beq A_d = 1+ \frac{C_d}{D_d} \; ; \; B_d = \frac{1}{D_d} \eeq
we will obtain 
\beq {\cal L} = \partial^{\mu} \phi \partial_{\mu} \phi \eeq
as desired. 

This, of course, is less than satisfactory. After all, the causal set is relativistic by the very setup, so one would expect relativistic covariance to appear naturally, while in the above situation if the coefficients $A_d$ and $B_d$ were not adjusted properly we would get non-covariant expression, where the non-covariance will manifest itself in $v$-dependence. On the other hand, in \cite{paper5} v-dependence is embraced with the explanation that $v$ is a function of $\phi$ and thus, overall, everything is covariant. The price, however, is that $v$ depends on $\phi$ in non-linear way, which causes a problem. On the other hand, in \cite{Proceedings} one gets rid of $v$-dependence, which allows $v$ to be random, which gets rid of non-linearity. In this respect one can claim that unnatural way of defining $A_d$ and $B_d$ is the price one can choose to pay in order to regain linearity. 

However, it turns out that this price is not sufficient in the case of non-linear $\phi$. In light of the fact that $\tau$ is finite, we can not assume exact linearity in a region of size $\tau$. In fact, since most of the values of $v$ are arbitrary close to the light cone, the Alexandrov sets corresponding to these $v$ "stretch out" arbitrary far coordinate-wise, which results in arbitrary large error in computting Lagrangian density for non-linear $\phi$. In order to address this issue, in \cite{Proceedings} some specific procedure was introduced through which, for any fixed $p$, one can select $v$ in a way that would minimize the fluctuations of $\phi$ in the interior of Alexandrov set $\alpha (exp_p (-\tau v/2), exp_p (\tau v/2))$ (thus, if there is at least one frame in which $\phi$ is approximately linear, $v$ will correspond to such a frame). This, of course, implies that $v$ is a function of both $p$ and $\phi$, just like it was in \cite{paper5}. The $\phi$ dependence is just as much non-linear in one case as in the other. Thus, we have established that \cite{Proceedings} is more "linear" theory than \cite{paper5} only in a special case where $\phi$ is linear. Since path integration takes us outside of that special case, sticking to \cite{Proceedings} doesn't adress linearity issues, nor does it allow us to do perturbation theory.  

This is merely a symptom of a far more general problem. In light of the non-compactness of Lorentz group, the Lorentzian neighborhood has infinitely large volume as well as infinitely large projections on coordinate axes. As a result, we were forced to select sub-neighborhood which has small volume, and small projection on at least one coordinate system. That sub-neighborhood defines a preferred frame, which is designated by a vector $v^{\mu}$ which is stationary in that frame (in the above cases the correspondence was established through exponential map; in general, it doesn't have to be that way, but still there would be \emph{some} kind of correspondence). In order for the choice of $v^{\mu}$ not to violate relativity, $v^{\mu}$ has to be a function of the behavior of the fields. That function is non-linear, which brings non-linearity into the theory. 

In the PST-like case described in this paper both problems can be reversed. Since the neigborhood in PST has finite size and finite projections on PST coordinates, both of which can be made arbitrary small, we no longer need to select sub-neighborhood. To put it another way, a finite PST neighborhood is a sub-neighborhood of infinite OST one, thus replacing OST with PST fulfilled our need of selecting sub-neighborhood. As stated in the analogy between translational and rotational covariance in the introduction, replacing OST with PST essentially allows us to violate relativity \emph{legally}. In this case, it provides a legal way of defining sub-neighborhood. Furthermore, in light of PST, $v^{\mu}$ becomes one of the variables, which means that it can no longer be viewed as a (non-linear) function of $\phi$. This, of course, removes one and only source of non-linearity of the theory.

We still retain the claim of \cite{paper5} that $v^{\mu}$ approximately coincides with the direction of gradient of $\phi$. But, since $\phi$ is now a function of $v^{\mu}$ and not the other way around, this is accomplished by imposing dynamics on $\phi$ rather than $v^{\mu}$. Thus, the issue of non-linearity of $v^{\mu}$ is avoided. An obvious question one might ask is that we might have just traded non-linearity of $v^{\mu}$ for non-linearity of $\phi$, by imposing non-linear constraint on behavior of $\phi$ with respect to rotation of $\partial^{\mu} \phi$ relative to $v^{\mu}$. This can be answered as follows: while it is true that imposing a restriction on $\phi$ is non-linear, if we expand $\phi$ as a Fourier series,
\beq \phi (x, v) = \int d^d k e^{ikx} \phi (k, v) \eeq
then the above restriction simply amounts to either significantly reducing coefficients of the harmonics $v(k, v)$ for which $k$ and $v$ have significant enough mismatch or throwing them away entirely. At the same time, what we do with each of the harmonics remains linear. Even if some harmonics have significantly reduced, or no, impact, we still can draw Feynmann diagrams, while remembering $v$ and $k$ dependence of the coefficients that we put before the propagator. Nevertheless, as will be shown in the section 4.3, the difference in coefficients will cause significant mismatch with well known results of quantum field theory, unless the situation is treated with great care. 

\noindent{\bf 4.2 Curvature effects}

The discussion in the previous section was assuming flat space time. One has to be a little more careful when curvature is introduced. If we parallel transport a vector around an arbitrarily large loop perpendicular to itself, the vector will be tilted by arbitrary large amount once we get to original point. So, if $\phi$ does not change as we perform such parallel transport, we would be forced to conclude that $\phi (x, v) = \phi (x, w)$, for all $v$ and $w$. Thus, $\phi$ would effectively become a function of $x$ alone, which would bring us right back to where we were before we introduced PST. In order to avoid this, we have to allow $\phi(x, v)$ to vary a little bit in the direction perpendicular to $v$. We simply have to impose an upper bound on that variation, which happens to be a very small number. This still means that we will have a problem in case that spacetime curvature is very high. For example, once the curvature is high enough, different wave modes can no longer be viewed as independent variables. This is related to the issue discussed in \cite{Davies} where it was shown that due to the global nature of the notion of a particle, such notion is not well defined in a curved spacetime. 

In \cite{Davies} this issue was addressed by hypothesizing asymptotically flat spacetime. I consider it unsatisfactory because if the spacetime is asymptotically flat, the non-flat region has a global "shape" that would break relativistic covariance. In fact, this is more cynical than the finite size of the universe postulated through big bang theory. In case of the latter, it was suggested by a lot of prominent names in cosmology that we should picturing ourselves to live on a four dimensional surface of a baloon that expands in higher dimensions; this means that it doesn't have any boundaries within the four dimensions. The issue of assimptotically flat behavior is much harder to treat since it is assumed to be found within our four dimensional world. It is possible to experiment with taking extra dimensions seriously and trying to take advantage of flatness in the region "outside of the baloon", but the success of this endeavor is questionable. Of course, from cosmological point of view it is \emph{possible} that our universe has well defined shape, regardless of what big bang theorists want us to believe. At the same time, the halmark of relativistic covariant theory is that it doesn't \emph{demand} that. In fact, as a local theory, it shouldn't demand anything at all on a larger scale. In other words, it has to be compatible with infinite universe, whether the universe is actually infinite or not. 

One might argue that in light of replacing OST with PST there is a way to somewhat remedy the situation. At least in the case of flat spacetime we can "slice" PST subsets with constant velocity coordinates in each slice, and claim that each slice has its own shape so that the "prefered direction" determined by the shape of that specific slice coincides with the direction of the common velocity of its elements. This, however, does not adress all the problems. First of all, as was just mentioned, we were considering global slices of constant velocity. This means that we had to assume flat spacetime in order for that slicing to be well defined. Consequently, this does not adress the curved spacetime scenario, which is what we were trying to adress in the first place. Secondly, even if we argue that rotational covariance is not violated, such boundaries of the universe still violate the translational one. Of course, one can argue that the presence of big bang and finite age of the universe violates translational covariance as well. But as was stated before, in order for the theory to be relativistic, neither big bang nor assimptotic behavior should be demanded by the theory.   

For these reasons, I hope to develop a theory where asimptotic spacetime is only a special case. At the same time, we will hypothesize that we live inside of that special case. We can then try to propose a theory that is well defined in general case. But this special case is the only scenario where the theory has to agree with a traditional quantum field theory in a curved spacetime. In particular, I claim that in a general case the presence of curvature prevents us from using principles of superposition, so the results can only be obtained numerically. But in the special case of assymptically flat spacetime, we can attempt to predict the behavior of particles in assymptotically flat region. That behavior will include "remembering" what happened in the curved spacetime region. The prediction of what would be "remembered" a billion years after the fact will be equated with the prediction of actual events. My hope is that decoherence phenomenon will help to define mechanism of this memory. This, however, is questionable since decoherence is designed to work for the case of superposition of harmonics, and curvature, as stated earlier, mixes otherwise separate hormonics of fields propagating on curved background, even though curvature is assumed to be non-fluctuating (which means that there is no gravitational field as such). The exploration of these questions is still up to future research. 

\noindent{\bf 4.3 The dynamics that limits unwanted degrees of freedom} 

In this section, we will take a brief look at different kinds of ways of limiting the variation of the field perpendicular to the direction of $v^{\mu}$. Let us start by exploring the ways \emph{not} to do it, and why they fail. The first thing one can think of is to add to the Lagrangian the derivative term in the direction perpendicular to $v^{\mu}$ with a very large coefficient, C:
\beq {\cal L} = m^2 \phi^2 + \Big(v^{\mu} \frac{\partial \phi}{\partial x^{\mu}} \Big)^2 + C^2 \Big(\Big(v^{\mu} \frac{\partial \phi}{\partial x^{\mu}} \Big)^2 -  \frac{\partial \phi}{\partial^{\mu} \phi} \frac{\partial \phi}{\partial_{\mu} \phi} \Big) \eeq 
If we expand $\phi$ as 
\beq \phi (x, v) = \int d^d k e^{ikx} \phi (k, v) \eeq
this would result in a very large variation of Lagrangian, whenever the difference between $v$ and $k$ is not too small, thus giving us hope that, in such situation, the nearby trajectories will cancel each other by interference. The propagator associated with this is
\beq \frac{1}{m^2 + (kv)^2 + C ((kv)^2 - k^2)} \eeq
where $kv=k^{\mu} v_{\mu}$ is a summation over OST indices. On the one hand, the propagator is a very small number, unless $v$ and $k$ are very close, as desired. On the other hand, however, due to the fact that $C$ is so large, even when $v$ and $k$ are, in fact, very close, the propagator might vary wildly. Roughly speaking, it amounts to a particle having substantially larger mass, as soon as $k$ and $v$ deviate from each other by very small amount. Furthermore, that "mass" variation is continuous. This, of course, contradicts everything we see in a lab, which means that this approach does not work.

The other idea is to multiply $C$ by an imaginary unit $i$. The motivation for doing that is that in the path integral it would give a factor $e^{i(iC)} = e^{-C}$, which is a very small number. But, if we write down the propagator, we will get 
\beq \frac{1}{m^2 + (kv)^2 + iC (k^2 - (kv)^2)} \eeq
This propagator mimics all the properties of the previous one: its value is very small unless $k$ and $p$ are very close to each other, but, unfortunately, it changes by very large amounts when they are. The reason $e^{i(iC)} = e^{-C}$ didn't solve anything is that, in the situation when $k$ and $p$ are very close, $e^{-C(k^2 - (kv)^2)}$ is no longer very small; thus, its variation due to large value of $C$ makes non-neglegeable difference. 

Thus, the key issue in both of the above cases is that we would like to have it both ways: on the one hand, we would like $C$ to be very large so we can force $v$ to be close to $k$, and, on the other hand, we don't want it to be too large, so that we can use the fact that $v$ and $k$ are close to claim that the difference between the two is unimportant.The problem is that formal multiplication by either $C$ or $iC$ doesn't allow us to have it both ways. 

The solution to this dilemma is to restrict the range of integration by hand instead of imposing extra terms into Lagrangian. One can argue that, from strictly mathematical point of view, we are already used to doing that in standard quantum field theory. In particular, we use this concept when we postulate ultraviolet or infrared cutoff in path integration.  In the situation at hand, we postulate that $(\partial \phi / \partial x^{\mu})(x^{\mu}, v^{\mu})$ is \emph{almost} parallel to $v^{\mu}$. In other words, the local space derivative of $\phi$, in a reference frame corresponding to a given element of PST, can not be larger than some small constant. In this case we can, in fact, "have it both ways". The restriction, being all-or-nothing by nature, does not have any effect at all as long as the value of the derivative is smaller than that constant; at the same time, it has the largest possible effect once it gets larger (even by a very small amount), which amounts to getting rid of unwanted terms entirely. 

A causal set version of the restriction of the spacelike derivative of $\phi$ can be defined by restricting our range of integration to $\phi \in {\cal F}_{\delta, \epsilon; n}$, where ${\cal F}_{\delta, \epsilon ; n}$ is defined as follows:

\textbf{Definition}: Let $\phi \colon S \rightarrow \mathbb{R}$ be a scalar field,  $\epsilon \in \mathbb{R}^+$, $\delta \in \mathbb{R}^+$ and $n \in \mathbb{N}^+$. Then $\phi$ is an element of ${\cal F}_{\delta, \epsilon, n}$ if and only if, whenever $t_p (q)$ and $v_{p;n} (q)$ are both less than $\delta^2$, while $r_p (q)$ is less than $\delta$, we always have $\vert \phi (q) - \phi (p) \vert < \epsilon \delta$.

In the above definition, the contrast between $\delta^2$ being an upper bound on the shifts of $t_p (q)$ and $v_{p; n} (q)$, and $\delta$ being an upper bound on the shift of $r_p (q)$ indicates that the direction of the shift is parallel to the direction of the radius, up to a small angle $\delta$ (which, for pure convenience of saving letters in the alphabet, happened to coincide with a small size of a neighborhood, although the two have nothing to do with each other). Now, since we are within $\delta$ neighborhood, the variation of the field is expected to be bounded by something first-order-small, or, in other words, something of the order $\delta$. Thus, the fact that it is bounded by $\epsilon \delta$ instead, which is second order small, indicates that the derivative of $\phi$ in the direction of the shift is first order small. As we have established, the direction of the shift is parallel to radial direction up to first order variations. Thus, the above definition implies that the radial derivative of $\phi$ is first order small as well. This is equivalent to saying that the OST projection of the derivative of $\phi$ is almost parallel to $t$ axis in a reference frame of $PST$ element of interest. This, in fact, is the ultimate intention of the above definition. 

Finally, we implement the restriction $\phi \in {\cal F}_{\delta, \epsilon, n}$ by defining the probability amplitude as the integral \emph{only} over such $\phi$:
\beq Z_{\delta, \epsilon, n} = ln \int_{\phi \in {\cal F}_{\delta, \epsilon, n}} [{\cal D} \phi] e^{iS(\phi)} \eeq  

\noindent{\bf 4.4 Kinetic, mass and interaction terms of the Lagrangian}

Now that we spent some time discussing ways of getting rid of unwanted degrees of freedom, we will now shift gears and describe the degrees of freedom that we do want. In particular, we will analyze the structure of non-interacting as well as interacting terms of the Lagrangian and the way it differs from the OST case.  

In light of the fact that the gradient points roughly in $v^{\mu}$ direction, the kinetic term of Lagrangian for a scalar field can be written as
\beq {\cal L} (\phi) (x, v) = \Big( v^{\mu} \frac{\partial \phi}{\partial x^{\mu}} \Big)^2 \eeq
where $\mu$ runs only over OST coordinates. The causal set version of this is 
\beq {\cal L} (\phi)(p) = \Big( \sum_{j=-n}^n \sum_{q \in G_j (p)} j(\phi (q) - \phi (p)) \Big)^2 \eeq
On the other hand, in order to allow for the momentum exchange during the interaction of two different fields, the continuum version of the Lagrangian density associated with interaction of $\phi$ and $\chi$ is given by
\beq {\cal L} (\phi)( x, v) = \int_{ v', v'' \in N_{\Lambda} (v)} \phi (x, v') \chi (x, v'') \eeq
Here, $N_{\Lambda} (v)$ is a very large neighborhood of $v$, whose size is determined by a very large number $\Lambda$, which is viewed as ultraviolet cutoff of path integration. As before, the reason we can afford having $\Lambda$ large is that there is no curvature in velocity coordinates. As in ordinary quantum field theory, the conservation of $k$ is a result of the integration of Fourier harmonics that gives $0$ unless frequencies match. The conservation of $v$ is the result of the constraint $v \approx k$. This means that if the above coupling violates conservation of $v$, only the second order terms survive; the latter form a vortex which allows for conservation of both $k$ and $v$.  

In the discrete case, due to the fact that $e \approx (x, v)$, $e' \approx (x,v')$ and $e'' \approx (x, v'')$ are completely different elements of a causal set, we can not assume that they share the same value of $x$ (which is one reason I just put approximation signs). Thus, $x$ has to be likewise replaced with $x'$ and $x''$ in the last two cases, where both $x'$ and $x''$ are constrained to some neighborhood, $n_{\delta} (x)$. However, we would like to say that while $N_{\Lambda} (v)$ is very large, $n_{\delta} (x)$ is very small. Since $x$ and $v$ are indistinguishable from each other, there is a single neighborhood, $N_{\Lambda, \delta} (e)$, which consists of $e'$ whose $x$-component differs only slightly from $e$, while $v$-component differs a lot (I decided to use $e$ for an element of causal set in this particular section in order not to confuse $p$ with momentum; I am greatful to \cite{newSorkin} for giving me an idea to use this particular letter):
\beq N_{\Lambda, \delta; n}(e) = \{ e' \vert \; max \; (\vert t_e (e') \vert, \vert r_e (e') \vert)  < \delta \wedge \vert v_{e;n} (e') \vert < \Lambda \} \eeq 
Here, $n$ is the constant taken from the section 2.6 which is used to define velocity $v_{e; n} (q)$. The causal set version of the interaction Lagrangian is
\beq {\cal L}^{(n)}_{\rm int} (\phi, \chi) (e) = \sum_{\{ e' , e'' \} \subset N_{\Lambda, \delta; n} (e) } \phi (e') \chi (e'') \eeq
This means that, in the approximation where $\Lambda$ is assumed to be infinite, the interaction term looks like $\phi (x) \chi (x)$ rather than $\phi (x, v) \chi (x, v)$. However, one still uses $\phi (x, v)$ when it comes to non-interacting kinetic term, which is what makes it possible to introduce contraction of $\partial_{\mu} \phi$ with $v^{\mu}$. Physically, this means that no momentum exchange occurs during the propagation of a particle, while it does occur at vertexes. However, if one chooses to interpret momentum exchange in terms of shift in $v^{\mu}$ between steps as opposed to coupling between different $v^{\mu}$, one would find that, in light of the nature of random walk in PST, there is a small momentum exchange even in case of free Lagrangian. This can be vaguely interpretted as "momentum exchange between the propagating particle and the elements of $S$", and this can be viewed as vacuum energy. This picture, however, is very vague since the lifetime of each element of $S$ is $0$ which doesn't allow them to "live" long enough to gain or lose energy. 

Now lets look at a mass term. Despite the fact that algebraically it looks like the interaction of $\phi$ with itself, in reality it does not involve any interaction at all, as it is part of a free propagator. In case of quantum field theory based on OST, this understanding doesn't affect the way we do algebra. In case of PST, however, the fact that no interaction occurs has to be explicitly incorporated in that we do not have a coupling $\phi (x, v) \phi (x, w)$ if $w^{\mu}$ is too different from $v^{\mu}$.  Thus, the mass term is given by 
\beq {\cal L}_{mass} (\phi)(x, v) = m^2 \phi^* (x, v) \phi (x, v) \eeq
and causal set version of this is simply
\beq {\cal L}_{mass} (\phi)(e) = m^2 \phi^* (e) \phi (e) \eeq 
However, self-interaction $\phi^4$ term continues to be defined in a way that allows for momentum exchange, since that is what we expect when we perform $\phi^4$ loop diagrams:
\beq {\cal L}_{\rm int} (\phi) (x, v) = \lambda \int_{v_i \in N_{\Lambda} (v)} \phi (x, v_1) \phi (x, v_2) \phi (x, v_3) \phi (x, v_4) \eeq
and the causal set version of this is 
\beq {\cal L}^{(n)}_{\rm int} (\phi) (p) = \lambda \sum_{p_i \in N_{\Lambda, \delta; n} (p)} \phi (e_1) \phi (e_2) \phi (e_3) \phi (e_4) \eeq

\noindent{\bf 4.5 Electromagnetic field}

We conclude this section by introducing electromagnetic field on a causal set. In order to be able to give stationary frame of a photon, we will postulate a small, but non-zero mass for an electromagnetic field. The same was done on p. 36 of \cite{Zee}. 

In case of continuous geometry, we define electromagnetic field to be a function $A \colon {\cal M}_{PST} \rightarrow  {\cal M}_{PST}$. The Lagrangian for an electromagnetic field is given by
\beq {\cal L} = \frac{1}{4} F^{\mu \nu} F_{\mu \nu} \eeq
Similarly to what was done for scalar field, its values are restricted in such a way that $A(x, v)$ varies very little in a direction perpendicular to $v^{\mu}$. If we define $t$ axis to be the direction of $v^{\mu}$, we can get rid of all of the space derivative terms of the Lagrangian, which leads us to   
\beq {\cal L} = - \frac{1}{2} \sum (\partial^0 A^k)^2 \eeq 
Since we already know that a vector is identified with PST element, the first thing that comes to mind is to view electromagnetic field as a scalor field on PST. This would also suggest a very appealing concept of unification of scalar and gauge field. Unfortunately, however, according to the model presented in this paper, PST only consists of timelike vectors, while the field $A^k$ is spacelike. This, too, might be something to re-think in the future research: after all, if the electromagnetic field was timelike it would have a natural physical meaning, namely phase shift of a moving particle that interacts with it, while spacelike electromagnetic field lacks that meaning. Unfortunately, at this point I don't see how to make that compatible with the wave front picture that I would like to subscribe to in the current paper. For this reason, I am forced to stick with an idea of spacelike electromagnetic field as far as this work is concerned, which does not allow me to view it as a scalar field on PST; although I still intend to come back with these other ideas in future research.

In light of this, I have two choices. One choice is to consider two PST elements, one pointing forward in time and the other pointing backwards in time, and claim that these two timelike directions add up to the spacelike one. The other approach is to view the elements of PST in terms of their projection onto OST and notice that a pair of spacelike separated elements (whether PST or OST) can determine a spacelike direction based on the geodesic that connects them. The latter has been an approach in OST-based theories of electromagnetic field in \cite{paper5} and \cite{Proceedings}. We notice that, even though the reasons are very different, in both cases we are forced to select two PST elements rather than just one. Thus, neither of these models would give us the unification of gauge and scalar fields that we desire. In light of this, we stick to the view proposed in \cite{paper5} and \cite{Proceedings} simply because it is easier to impliment. 

According to \cite{paper5} and \cite{Proceedings} an electromagnetic field was defined as a holonomy, $a \colon S \times S \rightarrow \mathbb{R}$, which represents the path integral of $A^{\mu}$ along the geodesic connecting a given pair of points. Strictly speaking, in case of a function $f \colon S \rightarrow \cal M$, the pullback of $A^{\mu}$ on $\cal M$ is given by
\beq (f^*A)(p,q) = \int_{\gamma (f(p), f(q))} g_{\mu \nu} A^{\mu} dx^{\nu} \eeq
where $\gamma (f(p), f(q))$ is a geodesic connecting $f(p)$ with $f(q)$ (see \cite{England}).  

Now, a causal set version of the restriction $A^0 \approx 0$ and  $\partial^k A^{\mu} \approx 0$ is $a \in {\cal A}^{(1)}_{\delta, \epsilon; n}$ and $a \in {\cal A}^{(2)}_{\delta, \epsilon; n}$ respectively. We would now like to rigorously define these. In order to enforce locality, we restrict ourselves to the neigbhborhood defined by $max(\vert t_p (q) \vert, \vert r_{p;n} (q) \vert , \vert v_{p;n} (q) \vert ) < \delta$. In order to speak of $A^0$, we have to speak of $a(q,s)$ where $q$ and $s$ are shifted in $t$ direction relative to each other. This can be enforced by saying that their $r$-shift is of the order of $\delta^2$. This would leave $t$ and $v$ as the primary directions of the shift. Thus, we would have a combination of $A^0$ and $A^v$ (we will come back to $A^v$ shortly). Now, in light of the smallness of the neighborhood, $a(q,s)$ is \emph{expected} to be of order $\delta$. So, in order to imply that $A^0 =0$ we have to say, instead, that it is bounded by $\epsilon \delta$. Now, apart from implying $A^0 =0$, this will also imply $A^v =0$. I consider this to be an important side benefit. After all, if such weren't the case, we would have been forced to view $A^{\mu}$ as an element of a tangent bundle to PST (or, equivalently, a tangent bundle to a tangent bundle to OST) rather than PST itself (or equivalently a tangent bundle to OST).  While, technically, such is still the case, due to the unwanted degrees of freedom being nearly $0$, $A^{\mu}$ can still be \emph{approximated} as an element of PST.  Thus, we obtain the following definition: 

\textbf{Definition}: Let $\delta$ and $\epsilon$ be reals, and let $n$ be an integer. Let $a \colon S \times S \rightarrow \mathbb{R}$ be a real valued function. That function is an element of ${\cal A}^{(1)}_{\delta, \epsilon; n}$ if and only if whenever $max(\vert t_p (q) \vert, \vert r_{p;n} (q) \vert , \vert v_{p;n} (q) \vert ) < \delta$ and $\vert t_p (q) - t_p (r) \vert < \epsilon \delta$, we have $\vert a(q, r) \vert < \epsilon^2$

Now we will move on to the definition of ${\cal A}^{(2)}_{\delta, \epsilon; n}$. This requires a notion of two different pairs of elements, $(p, q)$ and $(p', q')$ which define the directions of the holonomy, corresponding to two different pairs of points,  being "parallel" to each other. In light of the fact that $v$-component of $A$ is nearly zero, these four elements of PST, or, equivalently vectors in OST, are viewed as points in OST, and the only relevant direction is the line that connects them. 

However, it is still "easier" to define the parallelism then it would have been in the case of true OST. In light of the fact that we would like to focuse on the space component of the vector potential, we would like to assume that $p$, $q$, $p'$ and $q'$ share the same hyperplane and point in perpendicular direction to that hyperplane. This can be stated more precisely by postulating that both their time and velocity coordinates with respect to each other are $0$.

We can now view the imaginary line connecting $p$ and $q$ as a "projection" of a geodesic passing through some other vector, $r$, onto a spacelike plane perpendicular to the common direction of $p$, $q$, $p'$ and $q'$. In order to define the projection, we need to draw lines passing through $p$, $q$, $p'$ and $q'$ that are perpendicular to that common plane. This is easy to do, since these lines are nothing but geodesics passing through these four PST elements (which are viewed as OST vectors). 

Thus, when we say that $p$ is a projection of an element of a geodesic passing through $r$, we mean that there are vectors $p^*$ and $r^*$ on geodesics passing through $p$ and $r$ respectively such that $r_{p^*} (r^*)$ and $r_{r^*} (p^*)$ are both close to $0$. Likewise, $p'$ and $q'$ can be viewed as a projection of a geodesic passing through some other vector, $r'$. 

We can define these two geodesics to be parallel to each other by postulating that their velocities with respect to each other are very small. Finally, we can postulate that, as long as these geodesics are parallel, the segments are parallel as well. Furthermore, the segments have equal length as long as these geodesic segments have equal length, where the length can be defined in terms of $t$-coordinate along each geodesic. 

This brings us to the following definition:

\textbf{Definition}: Let $\delta$ be real and $n$ be an integer. Let $p$, $q$, $p'$ and $q'$ be four elements of $S$. We say that $(p, q)$ is \emph{parallel} to $(p', q')$ up to $(\delta, n)$ if the following conditions are met:

a) $max ( \vert r_p (q) \vert , \vert r_p (p') \vert ,\vert r_{p'} (q') \vert ) < \delta$

b) $max ( \vert t_p (q) \vert , \vert t_p (p') \vert ,\vert t_{p'} (q') \vert ) < \delta^2$

c) $max ( \vert v_{p; n} (q) \vert , \vert v_{p; n} (p') \vert ,\vert r_{p'; n} (q') \vert ) < \delta^2$

d) There are elements $r$, $r'$, $r_1^*$, $r_2^*$, $r_1'^*$, $r_2'^*$, $p^*$, $q^*$, $p'^*$ and $q^*$ such that 

(i) $\{ r_1 ^*, r_2^* \} \subset G(r)$, $\{ r_1'^*, r_2'^* \} \subset G(r')$, $p^* \in G(p)$, $q^* \in G(q)$, $p'^* \in G(p')$, and $q'^* \in G(q')$

(ii) $max (r_{p^*} (r_1^*), r_{q^*} (r_2^*), r_{p'^*} (r_1'^*), r_{q'^*} (r_2'^*)) < \delta^2$

(iii) $max (t_{p^*} (r_1^*), t_{q^*} (r_2^*), t_{p'^*} (r_1'^*), t_{q'^*} (r_2'^*)) < \delta^2$

(iv) $max (v_{p; n} (p'), v_{p'; n} (p)) < \delta^2$

(v) $\vert t_{r_1 ^*} (r_2 ^*) - t_{r_1'^*} ( r_2'^*) \vert < \delta^2$

In the above definition part a, which employs $\delta$ rather than $\delta^2$, doesn't say that anything is "small". Instead, it simply assures locality. Parts b and c say that $t$ and $v$-components are, in fact very close to each other, while $r$ component is not (since $\delta^2 << \delta$), thus implying that the vectors share the same hyperplane and all point perpendicular to that hyperplane. In part d, (i)-(iii) imply that the each pair of elements is a projection of a pair of elements on corresponding geodesic. Part (iv) says that these geodesics are parallel to each other, and finally part (v) says that the segments of geodesics whose projections we are looking at have equal length, thus implying equal distances between the elements in each pair. 

We are now ready to define ${\cal A}^{(2)}_{\delta, \epsilon; n}$. Basically, the key is to say that if $(p,q)$ is parallel to $(p', q')$ up to some order, the difference between $a(p,q)$ and $a(p',q')$ should be very small. Now, due to the fact that the size of a neighborhood is $\delta$, if the line $l$ connecting $p$ and $q$ was tilted by arbitrary angle to the line $l'$ connecting $p'$ and $q'$, then the difference between $a(p, q)$, and $a (p', q')$ would have still been of order $\delta$. Furthermore, again due to the small size of $\delta$-neighborhood, even if the space derivative of $A^{\mu}$ was non-zero, the continuity would still demand that the angle between the two lines is of the order $\delta$. This means that the difference between holonomies is of the order $\delta^2$. 

Now, in order for us to say that space derivative is, in fact, $0$ we should say that the difference between holonomies is smaller than expected. In other words, we have to say that the difference is of the order of $\delta^3$. But, as usual, $\delta$ will be the restriction on a domain, while $\epsilon$ on the range of the field. It will be understood, however, that the two small parameters are of the same order of magnitude. This means that restriction on variation of $a(p,q)$ is $\epsilon \delta^2$. This brings the following definition:

\textbf{Definition}: Let $a \colon S \times S \rightarrow \mathbb{R}$ be a real valued function on a set of pairs of elements of $S$. The function $a$ is an element of ${\cal A}^{(2)}_{\delta, \epsilon; n}$ if and only if whenever $(p, q)$ is parallel to $(p', q')$ up to $(n, \delta^3)$, we have $\vert a(p,q) - a(p', q') \vert < \epsilon \delta^2$. 

This defines our domain of integration:
\beq Z = \int_{{\cal A}^{(1)}_{\delta, \epsilon; n}
\cap {\cal A}^{(2)}_{\delta, \epsilon; n}} [{\cal D} a] e^{iS(a)} \eeq
Now, the only thing that is left is to define $S(a)$. In light of the previous discussion, the only thing we need is $\partial^0 A^k$. This can be simply computed by using $a(\tilde{p}, \tilde{q}) - a(p,q)$, where $\tilde{p} \in G_m (p)$ and $\tilde{q} \in G_m (q)$. If we like, we can average the different values of $n$. Thus, the electromagnetic Lagrangian is given by 
\beq {\cal L}_{p; n, m, \delta} (a) = \sum_{min(\vert t_p (q) \vert, \vert v_{p;n} (q) \vert) < \delta^2; \vert r_p (q) \vert = \delta} \; \sum_{\tilde{p} \in G_m (p), \tilde{q} \in G_m (q)} (a(\tilde{p}, \tilde{q}) - a(p, q))^2 \eeq 

\noindent{\bf 5. Conclusion}

As Sorkin mentioned in \cite{Sorkin}, causal sets are highly non-local due to light cone singularity. However, in the present work it is shown that the problem of light cone singularities is completely solved once the causal set is replaced with PST. Mathematically, this is due to the fact that there is nothing illegal in "neighborhood" on an OST violating translational invariance; thus, similarly, there is nothing illegal for a neighborhood on a PST to violate Lorentz covariance, either. Thus, every element of a PST, which is a vector in OST, can use its own direction as a "preferred" one when it comes to cutting off near-lightcone region.

Physically, light cone singularities that arise in the case of OST are the result of uncertainty principle, since, if it wasn't for that, we wouldn't have to integrate over all possible trajectories. Thus, instead of viewing each element of a set as a localized object with infinite position and zero momentum uncertainty, we viewed it as a wave packet where both uncertainties are finite. Thus, a position mean, together with momentum mean, of that wave packet determines the element of a PST that corresponds to it. 

The notion of well defined direction assigned to each element of PST allows to define the geometrical parameters easily and then generalize them to an arbitrary partial ordered set. Of course, an arbitrary causal set does not have a geometry, which means that the definition of these geometrical quantities is very formal. But, the fact that these quantities are defined, allows us to come up with well defined notion of what we mean by a causal set being PST-like in terms of local maps.

Furthermore, in case of a PST, a natural way of defining coordinates arises, without any assumption of the presence of continuum geometry. These coordinates allow a natural way of defining a "PST-like" causal set. While it could have been possible to use similar mapping in defining "manifold-like" causal set in OST case, the light cone singularities would imply that a causal set is only manifold-like in some frames and not others. Thus, if we are to demand that such frames be found, we would fundamentally set these frames up as "preferred frames" compared to others. On the other hand, in PST case it is possible that causal set is manifoldlike in \emph{every single} frame because the velocity is viewed as extra dimension and thus the "fast moving fields" while still undergoing the Lorentz contraction, do not "overlap" with slow moving ones in terms of PST (while they do overlap in terms of OST). 

Now, of course, local maps can similarly be introduced for more usual, OST-based concept of a causal set. But, in light of a light-cone singularities, a causal set can't possibly be manifold-like in these fast-moving frames. Thus, one is forced to constrain it to be manifold-like in some frames but not others, which violates relativity. 

In case of a the discretization of PST, on the other hand, the notion of a local region is already established without violation of relativity. Thus, it is possible to postulate PST-like property of every single local region, without postulating any preferred frame.  After all, in every single frame, the behavior of the vectors that move fast with respect to that frame is singular, while the behavior of vectors that move slowly is not. The latter, and not the former, is what is used in defining the geometry. 

Finally, the fields and Lagrangians were successfully introduced, in such a way that analytic computations can be done for causal sets with arbitrary many vectors, without relying on numeric work. The key of being able to do analytic computations is superposition principle, which allows one to expand the field into Fourier harmonics. In previous papers (\cite{Proceedings} and \cite{paper5}) non-linear procedure was used in finding a frame in which the variation of a given field is reasonable. That procedure was introduced in order to get rid of light cone singularities. The non-linear nature of the analysis of the behavior of the field is what prevented us from using superposition principle. In this paper, this does not need to be done since we are told from the start exactly what frame corresponds to what element of a causal set; the dynamics is designed in such a way that, as long as we stick to these frames, we are guaranteed to avoid singularities.  

At the same time the theory still has some weaknesses. As far as axiomatization of Chapter 3 goes, no proof was made to rigourously show that these axioms formally imply manifold strucutre. In this sense it is worse than \cite{Robb} where they did do all the proofs, for OST case. However, while \cite{Robb} was concerned about continuum case, I am concerned about discrete case. In light of this it would be several levels harder to make such proofs. While this is still a good topic for further research, it is possible that we should just do numeric simulations instead (which haven't been done either and thus, too, a topic of future research) and just see empirically if a given axiom system is complete enough.

As far as chapter 4 its main weakness is that I have limitted the range of integration to the functions where the direction of OST derivative approximately coincide with a direction of vectors corresponding to PST element. While I did show a qualitative argument that this procedure would generate Fourier transform, this argument is not explicit enough to be satisfactory. Again, I believe the main obstacle from doing more explicit work is the difficulty associated with describing random distribution. It is possible that in the future work I would have to consider some toy models, such as, for example, different regular lattice structures "moving" with different velocities, superimposed on each other, and see what happens. Nevertheless it is still interesting to explore how much analytic work can be done for the case of random Poisson distribution in light of the fact that at least one of the main obstacles, namely non-locality, was taken away.

Finally, the other weakness of Chapter 4 is the fact that the spirit of a constraint that the direction of PST element vaguely coincides with the OST derivative of the field is a vague equivalent of the notion of "particle". After all, a particle of fixed momentum, which is a wave, can be described in terms of that constraint. However, \cite{Davies} shows that particles can not be objectively defined in a curved spacetime. Equivalently, a parallel transport around the loop of a vector does not bring it back to the original one. I have dealt with this issue by emposing an upper bound on curvature which then allowed me to say that the deviation from that constraint is "small but finite", and that number is a function of curvature. Again, however, this is rather vague and I would like to see more explicitly how it works out in case of curved spacetime, in future research.

\end{document}